\pdfoutput=1

\documentclass[prd,twocolumn,lengthcheck,superscriptaddress,letterpaper,nofootinbib,nopreprintnumbers,showpacs]{revtex4-1}

\usepackage{color}
\usepackage{graphicx}
\usepackage{float}
\usepackage{amsmath}
\usepackage{acronym}
\usepackage{multirow}
\usepackage{svn-multi}
\usepackage{import}
\usepackage{hyperref}
\usepackage[normalem]{ulem}
\usepackage{placeins}
\renewcommand{\today}{\number\day\space\ifcase\month\or
  January\or February\or March\or April\or May\or June\or
  July\or August\or September\or October\or November\or December\fi
  \space\number\year}
\newcommand{\vLhat}{\mbox{\boldmath${\hat{L}}$}}
\newcommand{\vL}{\mbox{\boldmath${L}$}}
\newcommand{\vR}{\mbox{\boldmath${r}$}}
\newcommand{\vS}{\mbox{\boldmath${S}$}}

\newcommand{\Msun}{\ensuremath{\mathrm{M}_\odot}}

\newcommand{\TheEvent}{GW150914}

\newcommand{\macro}[1]{\textcolor{black}{#1}} 

\newcommand\OBSEVENTDATEMONTHDAYYEAR{\macro{September~14,~2015}}

\begin{document}

\title{An improved analysis of GW150914 using a fully spin-precessing waveform model}


\let\mymaketitle\maketitle
\let\myauthor\author
\let\myaffiliation\affiliation
\author{The LIGO Scientific Collaboration}
\author{The Virgo Collaboration}

\begin{abstract}
This paper presents updated estimates of source parameters for \TheEvent, a binary black-hole coalescence event detected by the Laser Interferometer Gravitational-wave Observatory (LIGO) on \OBSEVENTDATEMONTHDAYYEAR\ \cite{GW150914-DETECTION}. Reference\ \cite{GW150914-PARAMESTIM} presented parameter estimation of the source using a 13-dimensional, phenomenological precessing-spin model (precessing {IMRPhenom}) and a 11-dimensional nonprecessing effective-one-body (EOB) model calibrated to numerical-relativity simulations, which forces spin alignment (nonprecessing EOBNR). Here we present new results that include a 15-dimensional precessing-spin waveform model (precessing EOBNR) developed within the EOB formalism. We find good agreement with the parameters estimated previously \cite{GW150914-PARAMESTIM}, and we quote updated component masses of $35^{+5}_{-3}$\ \Msun\ and $30^{+3}_{-4}$\ \Msun (where errors correspond to 90\% symmetric credible intervals).
We also present slightly tighter constraints on the dimensionless spin magnitudes of the two black holes, with a primary spin estimate $<0.65$ and a secondary spin estimate $<0.75$ at 90\% probability.
Reference\ \cite{GW150914-PARAMESTIM} estimated the systematic parameter-extraction errors due to waveform-model uncertainty by combining the posterior probability densities of precessing {IMRPhenom} and nonprecessing EOBNR. Here
we find that the two precessing-spin models are in closer agreement, suggesting that these systematic errors are smaller than previously quoted.
\end{abstract}

\maketitle

\acrodef{BBH}[BBH]{binary black hole}
\acrodef{BNS}[BNS]{binary neutron star}
\acrodef{NS}[NS]{neutron star}
\acrodef{BHNS}[BHNS]{black hole--neutron star binaries}
\acrodef{PBH}[PBH]{primordial black hole binaries}
\acrodef{SNR}[SNR]{signal-to-noise ratio}
\acrodef{LIGO}[LIGO]{Laser Interferometer Gravitational-wave Observatory}
\acrodef{LHO}[LHO]{LIGO Hanford}
\acrodef{LLO}[LLO]{LIGO Livingston}
\acrodef{LSC}[LSC]{LIGO Scientific Collaboration}
\acrodef{CBC}[CBC]{compact binary coalescence}
\acrodef{GW}[GW]{gravitational wave}
\acrodef{FAR}[FAR]{false alarm rate}
\acrodef{FAP}[FAP]{false alarm probability}
\acrodef{IFO}[IFO]{interferometer}
\acrodef{BH}[BH]{black hole}
\acrodef{NR}[NR]{numerical relativity}
\acrodef{PN}[PN]{post-Newtonian}
\acrodef{EOB}[EOB]{effective-one-body}
\acrodef{PSD}[PSD]{power spectral density}
\acrodef{PDF}[PDF]{probability density function}
\acrodef{PE}[PE]{parameter-estimation}
\acrodef{LAL}[LAL]{LIGO Algorithm Library}
\acrodef{MaP}[MaP]{maximum a posteriori probability}
\acrodef{NR}[NR]{numerical relativity}

\section{Introduction}
The detection of the first gravitational-wave (GW) transient, \TheEvent\,, by the \ac{LIGO} on \OBSEVENTDATEMONTHDAYYEAR{} \cite{GW150914-DETECTION} marked the beginning of a new kind of astronomy, fundamentally different from electromagnetic or particle astronomy.
\TheEvent\ was analyzed using the most accurate signal models available at the time of observation, which were developed under the assumption that general relativity is the correct theory of gravity.
The analysis concluded that \TheEvent\ was generated by the coalescence of two black holes (BHs) of rest-frame masses $36^{+5}_{-4}\ \Msun$ and $29^{+4}_{-4}\ \Msun$, at a luminosity distance of $410^{+160}_{-180}\ \mathrm{Mpc}$ \cite{GW150914-PARAMESTIM}. Throughout this paper, we quote parameter estimates as the median of their posterior probability density, together with the width of the 90\% symmetric credible interval.

The GW signal emitted by a \ac{BBH} depends on 15 independent parameters: the BH masses, the BH spin vectors, the line of sight to the detector (parametrized by two angles), the sky location of the binary (parametrized by two angles), the polarization angle of the GW, the luminosity distance of the binary, and the time of arrival of the GW at the detector.
The task of extracting all 15 parameters from interferometric detector data relies on efficient Bayesian inference algorithms and on the availability of accurate theoretical predictions of the GW signal. State-of-the-art numerical-relativity (NR) simulations~\cite{Bruegmann:2006at,O'Shaughnessy:2012ay,Scheel:2014ina,
Chu:2015kft,Lousto:2015uwa,Szilagyi:2015rwa} 
can generate very accurate \ac{BBH} waveforms over a large region of parameter space; however, this region does not yet include (i) binary configurations that have both large dimensionless spins ($> 0.5$), exterme mass ratios ($> 1/3$) and many GW cycles ($\geq 40\mbox{--}60$), except for a few cases~\cite{Szilagyi:2015rwa,Kumar:2015tha,Lousto:2014ida}; nor does it include (ii) systems undergoing significant spin-induced precession of the orbital plane. In practice, since parameter estimation requires very many waveform evaluations that span a large region of the parameter space, a purely NR approach is unfeasible. Therefore, great effort has been devoted to building semi-analytical waveform models that are more efficient computationally than full-fledged NR, while still sufficiently accurate for the purposes of extracting unbiased physical information from the data.

The first parameter-estimation study of \TheEvent\ \cite{GW150914-PARAMESTIM} used two such models: 
an effective-one-body (EOB, \cite{Buonanno:1998gg, Buonanno:2000ef}) model that restricts spins to be aligned with the orbital angular momentum \cite{Taracchini:2013rva}, and a phenomenological model that includes spin-precession effects governed by four effective spin parameters \cite{Hannam:2013pra}.
Here we present updated parameter estimates using a \emph{fully spin-precessing} EOB model \cite{Pan:2013rra,BabaketalInPrep}, which is parametrized by the full set of \ac{BBH} properties listed above, including all six BH-spin degrees of freedom, and which reflects additional physical effects described in Sec.\ \ref{S:Model}. 
The inclusion of these effects motivates us to repeat the Bayesian analysis of \TheEvent\ with precessing EOB waveforms.
This model was not used in Ref.\ \cite{GW150914-PARAMESTIM} because it requires costly time-domain integration for each set of \ac{BBH} parameters; thus, not enough Monte Carlo samples had yet been collected by the time the study was finalized.

The main result of our analysis is that the two precessing models (phenomenological and EOB) are broadly consistent, more so than the precessing phenomenological and nonprecessing EOB models compared in Ref.\ \cite{GW150914-PARAMESTIM}.
In that study, the parameter estimates obtained with those two models were combined with equal weights to provide the fiducial values quoted in Ref.\ \cite{GW150914-DETECTION}, and they were differenced to characterize systematic errors due to waveform mismodeling.
Because the two precessing models yield closer results, we are now able to report smaller combined credible intervals, as well as smaller estimated systematic errors. Nevertheless, the combined medians cited as fiducial estimates in Ref.\ \cite{GW150914-DETECTION} change only slightly. 
In addition, we find that some of the \emph{intrinsic} parameters that affect \ac{BBH} evolution, such as the \emph{in-plane} combination of BH spins that governs precession, are constrained better using the precessing EOB model.

Because precessing-EOB waveforms are so computationally expensive to generate, we cannot match the number of Monte Carlo samples used in Ref.\ \cite{GW150914-PARAMESTIM}.
Thus, we carry out a careful statistical analysis to assess the errors of our summary statistics (posterior medians and credible intervals) due to the finite number of samples.
We apply the same analysis to the precessing phenomenological and nonprecessing EOB models, and to their combinations. Although finite-sample errors are a factor of a few larger for the precessing EOB model than for the other two, they remain much smaller than the credible intervals, so none of our conclusions are affected.
Last, as a further test on the accuracy and consistency of the two precessing models, we use them to estimate the known parameters of a \TheEvent-like NR waveform injected into LIGO data. The resulting posteriors are similar to those found for \TheEvent.

This article is organized as follows. In Sec.\ \ref{S:Model} we discuss the modeling of spin effects in the \ac{BBH} 
waveforms used in this paper. In Sec.\ \ref{S:Description} we describe our analysis. We present our results in Sec.\ \ref{S:Results}, and our conclusions in Sec.\ \ref{S:Conclusion}. Throughout the article  we adopt geometrized units, with $G = c = 1$.

\section{Modeling orbital precession in BBH waveform models}
\label{S:Model}

Astrophysical stellar-mass BHs are known to possess significant intrinsic spins, which can engender large effects in the late phase of \ac{BBH} coalescences: they affect the evolution of orbital frequency, and (if the BH spins are not aligned with the orbital angular momentum) they induce the precession of the orbital plane, modulating the fundamental chirping structure of emitted GWs in a manner dependent on the relative angular geometry of binary and observatory \cite{S6pepaper}. While measuring BH spins is interesting in its own right, the degree of their alignment and the resulting degree of precession hold precious clues to the astrophysical origin of stellar-mass \ac{BBH}s \cite{GW150914-ASTRO}: aligned spins suggest that the two BHs were born from an undisturbed binary star in which both components successively collapsed to BHs; nonaligned spins point to an origin from capture events and three-body interactions in dense stellar environments.

Clearly, the accurate modeling of BH-spin effects is crucial to \ac{BBH} parameter-estimation studies. Now, even state-of-the-art semianalytical waveform models still rely on a set of approximations that necessarily limit their accuracy. These include finite PN order, calibration to a limited number of NR simulations, rotation to precessing frames, and more. Thus, being able to compare parameter estimates performed with different waveform models, derived under different assumptions and approximations (e.g., in time- vs.\ frequency-domain formulations), becomes desirable to assess the systematic biases due to waveform mismodeling. While observing consistent results does not guarantee the absence of systematics (after all, multiple models could be wrong in the same way), the fact that we do not observe inconsistencies does increase our confidence in the models.

Such a comparison was performed in the original parameter-estimation study of \TheEvent\ \cite{GW150914-PARAMESTIM}, showing consistency between the precessing phenomenological model and the aligned-spin EOBNR model. This result matched the finding that the BH spins were approximately aligned in \TheEvent, or that precession effects were too weak to be detected, because of the small number of GW cycles and of the (putative) face-on/face-off presentation of the binary. Nevertheless, it may be argued that the conclusion of consistency remained suspect, because only one model in the analysis carried information about the effects of precession; conversely, the estimates of mismodeling systematics performed in Ref.\ \cite{GW150914-PARAMESTIM} were likely increased by the fact that the nonprecessing model would be biased by what little precession may be present in the signal.

The analysis presented in this article, which relies on two precessing-spin waveform families, removes both limitations, and sets up a more robust framework to assess systematic biases in future detections where spin effects play a larger role. In the rest of this section, we discuss the features and formulation of the fully precessing EOBNR model. The reader not interested in these technical details (and in the Bayesian-inference setup of Sec.\ \ref{S:Description}) may proceed directly to Sec.\ \ref{S:Results}.

The precessing EOBNR model (henceforth, ``precessing EOBNR'') used here describes inspiral--merger--ringdown (IMR) waveforms for coalescing, quasicircular BH binaries with mass ratio $0.01\leq q\equiv m_2/m_1 \leq 1$, dimensionless BH spin magnitudes $0\leq \chi_{1,2}\equiv |\vS_{1,2}|/m^2_{1,2} \leq 0.99$, and arbitrary BH spin orientations.\footnote{In \ac{LAL}, as well as in technical publications, the 
precessing EOBNR model that we use is called \texttt{SEOBNRv3}.} We denote with $m_{1,2}$ the masses of the component objects in the binary, and with $\vS_{1,2}$ their spin vectors.

The fundamental idea of EOB models consists in mapping the conservative dynamics of a binary to that of a spinning particle that moves in a deformed Kerr spacetime~\cite{Buonanno:1998gg, Buonanno:2000ef,Damour:2001tu,Buonanno:2005xu,Damour:2008qf,Barausse:2009xi,Barausse:2011ys,Damour:2014sva}, where the magnitude of the deformation is proportional to the mass ratio of the binary. This mapping can be seen as a resummation of post-Newtonian (PN) formulas~\cite{blanchet:2014} with the aim of extending their validity to the strong-field regime. As for dissipative effects, EOB models equate the loss of energy to the GW luminosity, which is expressed as a sum of squared amplitudes of the multipolar waveform modes. In the nonprecessing limit, the inspiral--plunge waveform modes are themselves resummations of PN expressions~\cite{Damour:2007xr,Damour:2008gu,Pan:2010hz}, and are functionals of the orbital dynamics. The ringdown signal is described by a linear superposition of the quasinormal modes~\cite{Vishveshwara:1970zz,Press:1971wr,Chandrasekhar:1975zza} of the remnant BH. 

EOB models can be tuned to NR by introducing adjustable parameters at high, unknown PN orders. For the precessing EOB model used in this work, the relevant calibration to NR was carried out in Ref.~\cite{Taracchini:2013rva} against 38 NR simulations of nonprecessing-spins systems 
from Ref.~\cite{Mroue:2013xna}, with mass ratios up to 1/8 and spin magnitudes up to almost extremal for equal-mass BBHs and up to 0.5 for unequal-mass BBHs. 

Furthermore, information from inspiral, merger and ringdown waveforms in the test-particle limit 
were also included in the EOBNR model~\cite{Barausse:2011kb,Taracchini:2014zpa}. Prescriptions for the onset and spectrum of ringdown for precessing BBHs were first given in Ref.~\cite{Pan:2013rra}, and significantly improved in Ref.~\cite{BabaketalInPrep}. 

In the model, the BH spin vectors precess according to 
\begin{equation}
\frac{\mathrm{d}\vS_{1,2}}{\mathrm{d}t} = \frac{\partial H_{\mathrm{EOB}}}{\partial \vS_{1,2}} \times \vS_{1,2}\,;
\end{equation}
when the BH spins are oriented generically, the orbital plane precesses with respect to an inertial observer. The orientation of the orbital plane is tracked by the Newtonian orbital angular momentum $\vL_{\rm N} \equiv \mu\, \vR \times \dot{\vR}$, where $\mu \equiv m_1 m_2/(m_1+m_2)$ and $\vR$ is the relative BH separation. One defines a (noninertial) precessing frame whose $z$-axis is aligned with $\vL_{\rm N}(t)$, and whose $x$- and $y$-axes obey the mimimum-rotation prescription of Ref.~\cite{buonanno:2004yd,Boyle:2011gg}. In this frame, the waveform amplitude and phase modulations induced by precession are minimized, as pointed out in several studies \cite{buonanno:2004yd,Schmidt:2010it, O'Shaughnessy:2011fx, Boyle:2011gg, Schmidt:2012rh}.

Thus, the construction of a precessing EOB waveform consists of the following steps: (i) compute orbital dynamics numerically, by solving Hamilton's equation for the EOB Hamiltonian, subject to energy loss, up until the light ring (or photon orbit) crossing; (ii) generate inspiral--plunge waveforms in the precessing frame as if the system were not precessing~\cite{Taracchini:2013rva}; (iii) rotate the waveforms to the inertial frame aligned with the direction of the remnant spin; (iv) generate the ringdown signal, and connect it smoothly to the inspiral--plunge signal; (v) rotate the waveforms to the inertial frame of the observer.

A phenomenological precessing-spins IMR model (henceforth, ``precessing IMRPhenom'') was proposed in Refs.\ \cite{Schmidt:2014iyl,Hannam:2013oca}.\footnote{In \ac{LAL} this precessing model is called \texttt{IMRPhenomPv2}.} These waveforms are generated in the frequency domain by rotating nonprecessing phenomenological waveforms~\cite{khan:2015jqa} from a precessing frame to the inertial frame of the observer, according to PN formulas that describe precession in terms of Euler angles. The underlying nonprecessing waveforms depend on the BH masses and on the two projections of the spins on the Newtonian angular momentum, with the spin of the BH formed through merger adjusted to also take into account the effect of the in-plane spin components. The influence of the in-plane spin components on the precession is modeled with a single spin parameter (a function of the two BH spins), and depends also on the initial phase of the binary in the orbital plane. Thus, this model has only four independent parameters to describe the six spin degrees of freedom, which is justified by the analysis of dominant spin effects performed in Ref.~\cite{Schmidt:2014iyl}.

While both precessing EOBNR and IMRPhenom models describe spin effects, there are important differences in how they account for precession, which is the main focus of this paper.
\begin{enumerate}
\item In precessing IMRPhenom, the precessing-frame inspiral--plunge waveforms are strictly nonprecessing waveforms, while for precessing EOBNR some precessional effects are included (such as spin--spin frequency and amplitude modulations), since the orbital dynamics that enters the nonprecessing expressions for the GW modes is fully precessing.
\item The precessing EOBNR merger--ringdown signal is generated in the inertial frame oriented along the total angular momentum of the remnant---the very frame where quasinormal mode frequencies are computed in BH perturbation theory. By contrast, precessing IMRPhenom generates the merger--ringdown signal directly in the precessing frame.
\item The IMRPhenom precessing-frame waveforms contain only the dominant $(2,\pm2)$ modes, while precessing EOBNR includes also $(2,\pm1)$ modes in the precessing frame, although these are not calibrated to NR.
\item In IMRPhenom, the frequency-domain rotation of the GW modes from the precessing frame to the inertial frame is based on approximate formulas (i.e., on the stationary-phase approximation), while precessing EOBNR computes the rotations fully in the time domain, where the formulas are straightforward.
\item In precessing IMRPhenom, the frequency-domain formulas for the Euler angles that parametrize the precession of the orbital plane with respect to a fixed inertial frame involve several approximations: 
in-plane spin components are orbit averaged;
the magnitude of the orbital angular momentum is approximated by its 2PN nonspinning expression;
the evolution of frequency is approximated as adiabatic;
and the PN formulas that regulate the behavior of the Euler angles at high frequencies are resummed partially.
By contrast, precessing EOBNR defines these Euler angles on the basis of the completely general motion of $\vL_{\rm N}(t)$; this motion is a direct consequence of the EOB dynamics, and as such it is sensitive to the full precessional dynamics of the six spin components.
\end{enumerate}

\textit{A priori}, it is not obvious that these approximations will not impact parameter estimation for a generic BBH.
However, as far as GW150914 is concerned, Ref.\ \cite{GW150914-PARAMESTIM} showed broadly consistent results between a precessing and a nonprecessing model; \textit{a fortiori} we should expect similar results between two precessing models.
Indeed, the \TheEvent\ binary is most likely face-off or face-on with respect to the line of sight to the detector, and the component masses are almost equal \cite{GW150914-PARAMESTIM}: both conditions imply that subdominant modes play a minor role.

The nonprecessing models that underlie both approximants were tested against a large catalog of NR simulations \cite{Taracchini:2013rva,khan:2015jqa,Kumar:2016dhh}, finding a high degree of accuracy in the GW150914 parameter region. However, it is important to bare in mind that these waveform models can differ from NR outside the region in which they were calibrated and they do not account for all possible physical effects that are relevant to generic BBHs, such us higher-order modes. Finally, neither of the two precessing models has been calibrated to any precessing NR simulation. Thus, we cannot exclude that current precessing models are affected by systematics.

Since the generation of precessing EOBNR waveforms (at least in the current implementation in LAL) is a rather time-consuming process,\footnote{Generating a nonspinning, equal-mass binary black-hole waveform for a total mass of 70~\Msun\ from a starting frequency of 20~Hz is about a factor of 20 slower for precessing EOBNR than for precessing IMRPhenom.} when carrying out parameter-estimation studies with this template family, we introduce a time-saving approximation at the level of the likelihood function. Namely, we marginalize over the arrival time and phase of the signal as if the waveforms contained only $(2,\pm2)$ inertial-frame modes, since in that case the marginalization can be performed analytically.
We have determined that the impact of this approximation is negligible by conducting a partial parameter-estimation study where we do not marginalize over the arrival time and phase. We can understand this physically for GW150914 because in a nearly face-on/face-off binary the $(2,\pm1)$ observer-frame modes are significantly sub-dominant compared to $(2,\pm2)$ modes.\footnote{By construction, the precessing IMRPhenom inertial-frame polarizations $h_{+,\times}$ depend on the arrival time and phase exactly as they would in a model that includes only $(2,\pm2)$ inertial-frame modes. Thus, although precessing IMRPhenom does include $(2,\pm1)$ inertial-frame modes, the analytical marginalization that we just discussed is exact.}

\section{Bayesian Inference Analysis}
\label{S:Description}

For each waveform model under consideration, we estimate the posterior probability density \cite{Bayes:1793, Jaynes:2003} for the BBH parameters, following the prescriptions of Ref.\ \cite{GW150914-PARAMESTIM}. To wit, we use the \ac{LAL} implementation of parallel-tempering Markov Chain Monte Carlo and nested sampling \cite{veitch:2014wba} to sample the posterior density $p(\boldsymbol{\vartheta}|\text{model, data})$ as a function of the parameter vector $\boldsymbol{\vartheta}$:
\begin{equation}
p(\boldsymbol{\vartheta}|\text{model, data}) \propto 
\mathcal{L}(\text{data}|\boldsymbol{\vartheta}) \times p(\boldsymbol{\vartheta}).
\end{equation}
To obtain the likelihood $\mathcal{L}(\text{data}|\boldsymbol{\vartheta})$, we first generate the GW polarizations $h_+(\boldsymbol{\vartheta}_\mathrm{intrinsic})$ and $h_{\times}(\boldsymbol{\vartheta}_\mathrm{intrinsic})$ according to the waveform model. We then combine the polarizations into the LIGO detector responses $h_{1,2}$ by way of the detector antenna patterns:
\begin{eqnarray}
h_k(\boldsymbol{\vartheta}) &=& h_{+}(\boldsymbol{\vartheta}_\mathrm{intrinsic})\,F^{(+)}_k(\boldsymbol{\vartheta}_\mathrm{extrinsic}) \nonumber \\
&& +h_{\times}(\boldsymbol{\vartheta}_\mathrm{intrinsic})\,F^{(\times)}_k(\boldsymbol{\vartheta}_\mathrm{extrinsic})\,.
\end{eqnarray}
Finally, we compute the likelihood as the sampling distribution of the residuals (i.e., the detector data $d_k$ minus the GW response $h_k(\boldsymbol{\vartheta})$), under the assumption that these are distributed as Gaussian noise characterized by the \ac{PSD} of nearby data \cite{veitch:2014wba}:
\begin{equation}
\mathcal{L}(\text{data} | \boldsymbol{\vartheta})\propto \exp\left[-\frac{1}{2} \sum_{k=1,2} \left\langle h_k(\boldsymbol{\vartheta}) - d_k \middle| h_k(\boldsymbol{\vartheta}) - d_k\right\rangle\right],
\end{equation}
where $\langle\cdot|\cdot\rangle$ denotes the noise-weighted inner product~\cite{Cutler:1994ys}.

The prior probability density $p(\boldsymbol{\vartheta})$ follows the choices of Ref.~\cite{GW150914-PARAMESTIM}.
In particular, we assume uniform mass priors $m_{1,2}\in [10, 80]~\Msun$, with the constraint $m_2 \le m_1$, and uniform spin-amplitude priors $a_{1,2} = |\vS_{1,2}|/m^2_{1,2}\in [0, 1]$, with spin directions distributed uniformly on the two-sphere; and we assume that sources are distributed uniformly in Euclidian volume, with their orbital orientation distributed uniformly on the two-sphere.
All the binary parameters that evolve during the inspiral (such as \emph{tilt} angles between the 
spins and the orbital angular momentum, $\theta_{LS_{1,2}}$) are defined at a reference GW frequency $f_\mathrm{ref} = 20$ Hz.

Following \cite{GW150914-PARAMESTIM}, we marginalize over the uncertainty in the calibration of LIGO data~\cite{GW150914-CALIBRATION}. This broadens the posteriors but reduces calibration biases. For the precessing EOBNR analysis, we also marginalize over the time of arrival and reference phase of the GW signal, following the prescription of Ref.~\cite{margphi}.

To assess whether the data is \emph{informative} with regard to a source parameter (i.e., where it \emph{updates} the prior significantly), we perform a Kolmogorov--Smirnov (KS) test.
Given an empirical distribution (in our case, the Monte Carlo posterior samples) and a probability distribution (in our case, the prior), the KS test measures the maximum deviation between the two cumulative distributions and associates a $p$-value to that: for samples generated from the probability distribution against which the test is performed, one expects a $p$-value around 0.5; $p$-values smaller than 0.05 indicate that the samples come from a different probability distribution with a high level of significance. The outcomes of our KS tests are only statements about how much the posteriors deviate from the respective priors, but they do not tell us anything about the astrophysical relevance of 90\% credible intervals.

\section{Results}
\label{S:Results}

\begin{table*}
 \caption{\label{T:results}Median values of source parameters of GW150914 as estimated with the two precessing waveform models, and with an equal-weight average of posteriors (in the ``Overall" column). The models are described in the text. Subscripts and superscripts indicate the range of the symmetric 90\% credible intervals. When useful, we quote 90\% credible bounds.}
 \begin{ruledtabular}
\begin{tabular}{l c c c}
& precessing EOBNR & precessing IMRPhenom & Overall\\
\hline
Detector-frame total mass $M/\Msun$ & $71.6^{+4.3}_{-4.1}$ & \
$70.9^{+4.0}_{-3.9}$ & $71.3^{+4.3}_{-4.1}$ \\
Detector-frame chirp mass $\mathcal{M}/\Msun$ & $30.9^{+2.0}_{-1.9}$ \
& $30.6^{+1.8}_{-1.8}$ & $30.8^{+1.9}_{-1.8}$ \\
Detector-frame primary mass $m_1/\Msun$ & $38.9^{+5.1}_{-3.7}$ & \
$38.5^{+5.6}_{-3.6}$ & $38.7^{+5.3}_{-3.7}$ \\
Detector-frame secondary mass $m_2/\Msun$ & $32.7^{+3.6}_{-4.8}$ & \
$32.2^{+3.6}_{-4.8}$ & $32.5^{+3.7}_{-4.8}$ \\
Detector-frame final mass $M_\mathrm{f}/\Msun$ & $68.3^{+3.8}_{-3.7}$ \
& $67.6^{+3.6}_{-3.5}$ & $68.0^{+3.8}_{-3.6}$ \\
\\
Source-frame total mass $M^{\mathrm{source}}/\Msun$ & \
$65.6^{+4.1}_{-3.8}$ & $65.0^{+4.0}_{-3.6}$ & $65.3^{+4.1}_{-3.7}$ \\
Source-frame chirp mass $\mathcal{M}^{\mathrm{source}}/\Msun$ & \
$28.3^{+1.8}_{-1.7}$ & $28.1^{+1.7}_{-1.6}$ & $28.2^{+1.8}_{-1.7}$ \\
Source-frame primary mass $m_1^{\mathrm{source}}/\Msun$ & \
$35.6^{+4.8}_{-3.4}$ & $35.3^{+5.2}_{-3.4}$ & $35.4^{+5.0}_{-3.4}$ \\
Source-frame secondary mass $m_2^{\mathrm{source}}/\Msun$ & \
$30.0^{+3.3}_{-4.4}$ & $29.6^{+3.3}_{-4.3}$ & $29.8^{+3.3}_{-4.3}$ \\
Source-frame final mass $M_\mathrm{f}^{\mathrm{source}}/\Msun$ & \
$62.5^{+3.7}_{-3.4}$ & $62.0^{+3.7}_{-3.3}$ & $62.2^{+3.7}_{-3.4}$ \\
\\
Mass ratio $q$ & $0.84^{+0.14}_{-0.20}$ & $0.84^{+0.14}_{-0.20}$ & \
$0.84^{+0.14}_{-0.20}$ \\
\\
Effective inspiral spin parameter $\chi_\mathrm{eff}$ \
&$-0.02^{+0.14}_{-0.16}$ & $-0.05^{+0.13}_{-0.15}$ & \
$-0.04^{+0.14}_{-0.16}$ \\
Effective precession spin parameter $\chi_\mathrm{p}$ \
&$0.28^{+0.38}_{-0.21}$ & $0.35^{+0.45}_{-0.27}$ & \
$0.31^{+0.44}_{-0.23}$ \\
Dimensionless primary spin magnitude $a_1$ &$0.22^{+0.43}_{-0.20}$ & \
$0.32^{+0.53}_{-0.29}$ & $0.26^{+0.52}_{-0.24}$ \\
Dimensionless secondary spin magnitude $a_2$ &$0.29^{+0.52}_{-0.27}$ \
& $0.34^{+0.54}_{-0.31}$ & $0.32^{+0.54}_{-0.29}$ \\
Final spin $a_\mathrm{f}$ &$0.68^{+0.05}_{-0.05}$ & \
$0.68^{+0.06}_{-0.06}$ & $0.68^{+0.05}_{-0.06}$ \\
\\
Luminosity distance $D_\mathrm{L}/\mathrm{Mpc}$ &$440^{+160}_{-180}$ \
& $440^{+150}_{-180}$ & $440^{+160}_{-180}$ \\
Source redshift $z$ &$0.094^{+0.032}_{-0.037}$ & \
$0.093^{+0.029}_{-0.036}$ & $0.093^{+0.030}_{-0.036}$ \\
\hline
Upper bound on primary spin magnitude $a_1$ & 0.54 & 0.74 & 0.65\\
Upper bound on secondary spin magnitude $a_2$ & 0.70 & 0.78 &0.75\\
Lower bound on mass ratio $q$ & 0.69 & 0.68 & 0.68
\end{tabular}
\end{ruledtabular}
\label{tab:parametersOld}
\end{table*}

\begin{figure*}
\caption{Comparing nonprecessing EOBNR (light yellow, top),
precessing IMRPhenom (light blue, middle),
and precessing EOBNR (light red, bottom) 
90\% credible intervals for select GW150914 source parameters. The darker intervals represent error estimates for (from left to right) the 5\%, 50\% and 95\% quantiles, estimated by Bayesian bootstrapping.}
\begin{tabular}{rlrl}
\includegraphics[width=0.7\columnwidth]{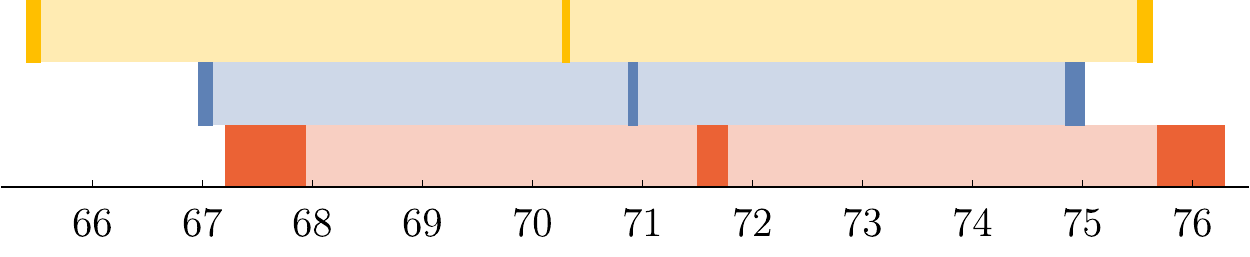}&$M/\textrm{M}_{\odot}$&\includegraphics[width=0.7\columnwidth]{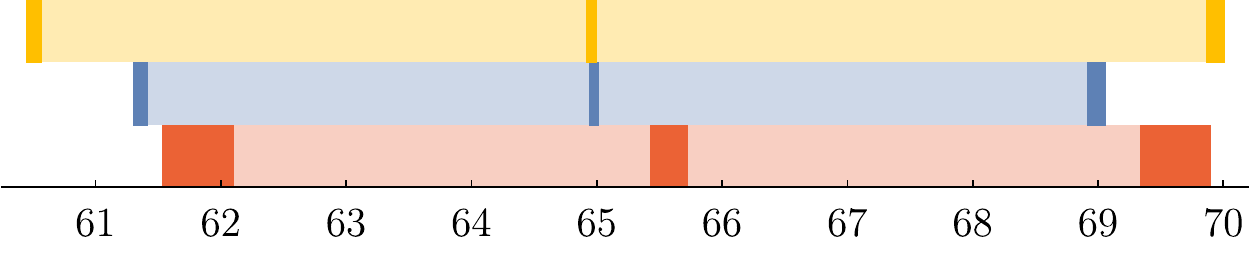}&$M^{\textrm{source}}/\textrm{M}_{\odot}$\\
\includegraphics[width=0.7\columnwidth]{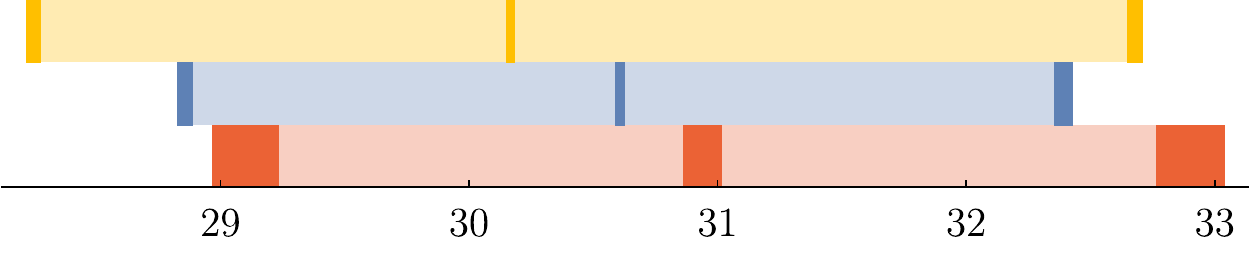}&$\mathcal{M}/\textrm{M}_{\odot}$&\includegraphics[width=0.7\columnwidth]{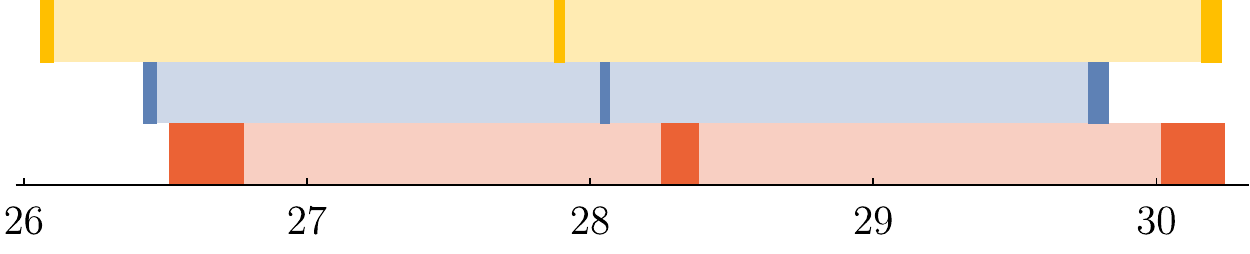}&$\mathcal{M}^{\textrm{source}}/\textrm{M}_{\odot}$\\
\includegraphics[width=0.7\columnwidth]{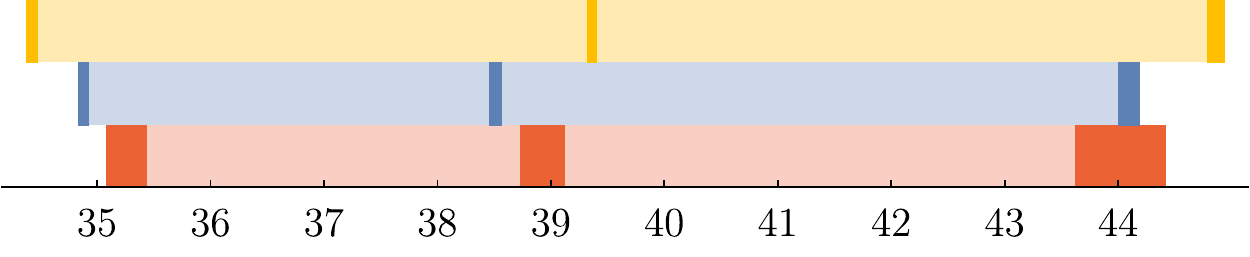}&$m_1/\textrm{M}_{\odot}$&\includegraphics[width=0.7\columnwidth]{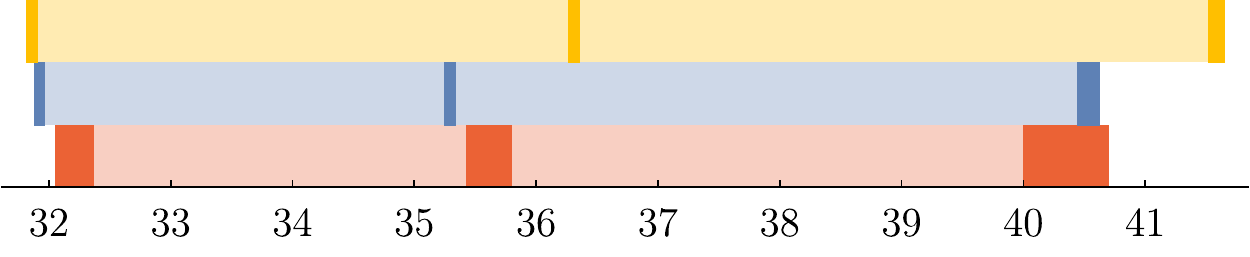}&$m_1^{\textrm{source}}/\textrm{M}_{\odot}$\\
\includegraphics[width=0.7\columnwidth]{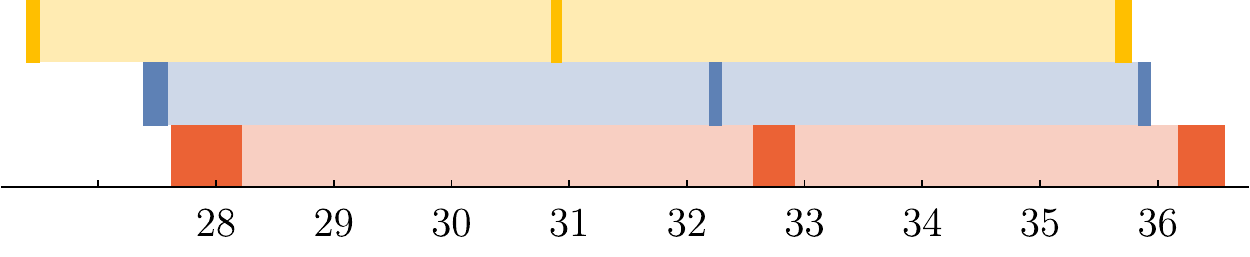}&$m_2/\textrm{M}_{\odot}$&\includegraphics[width=0.7\columnwidth]{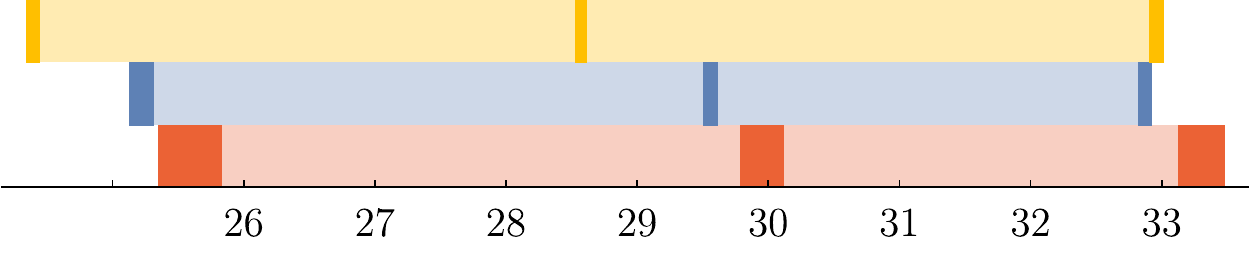}&$m_2^{\textrm{source}}/\textrm{M}_{\odot}$\\
\includegraphics[width=0.7\columnwidth]{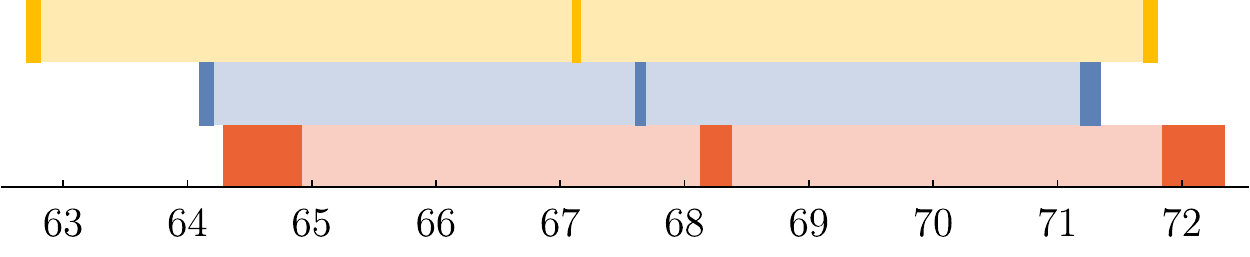}&$M_{\textrm{f}}/\textrm{M}_{\odot}$&
\includegraphics[width=0.7\columnwidth]{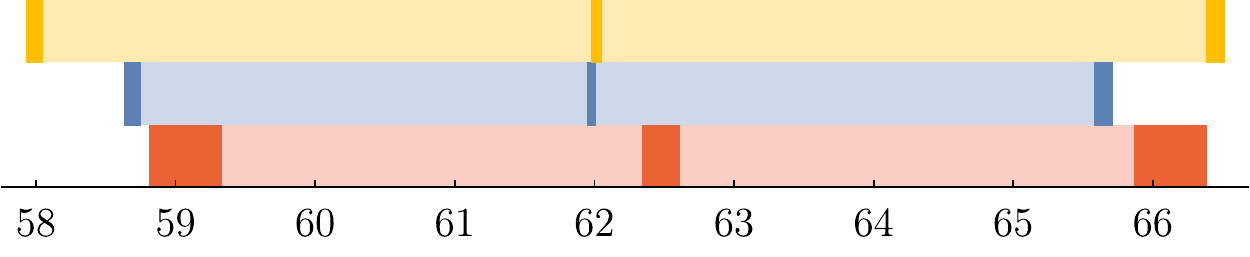}&$M_{\textrm{f}}^{\textrm{source}}/\textrm{M}_{\odot}$\\
\includegraphics[width=0.7\columnwidth]{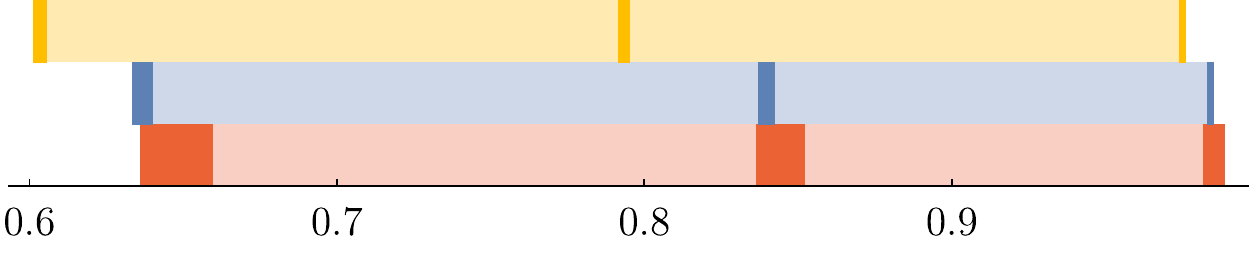} &$q$&&\\
\includegraphics[width=0.7\columnwidth]{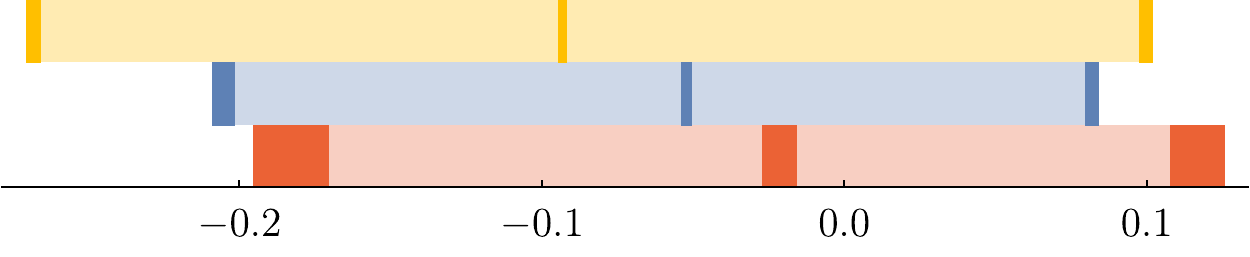}&$\chi_{\textrm{eff}}$&\includegraphics[width=0.7\columnwidth]{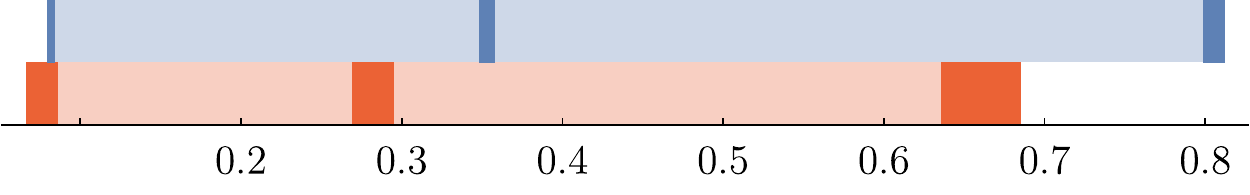}&$\chi_{\textrm{p}}$\\
\includegraphics[width=0.7\columnwidth]{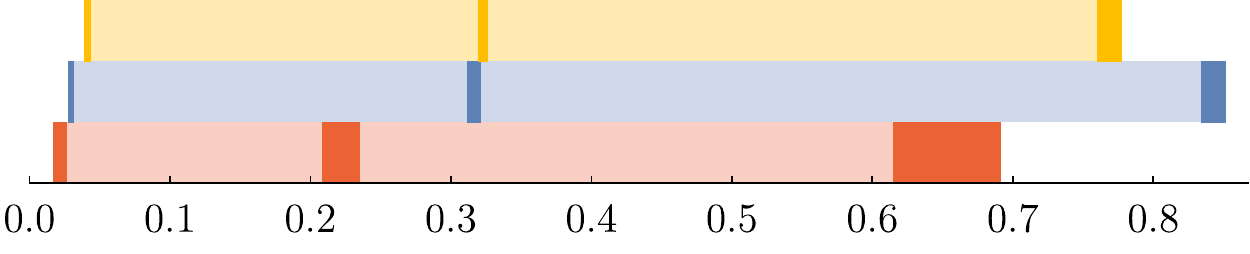}&$a_1$&\includegraphics[width=0.7\columnwidth]{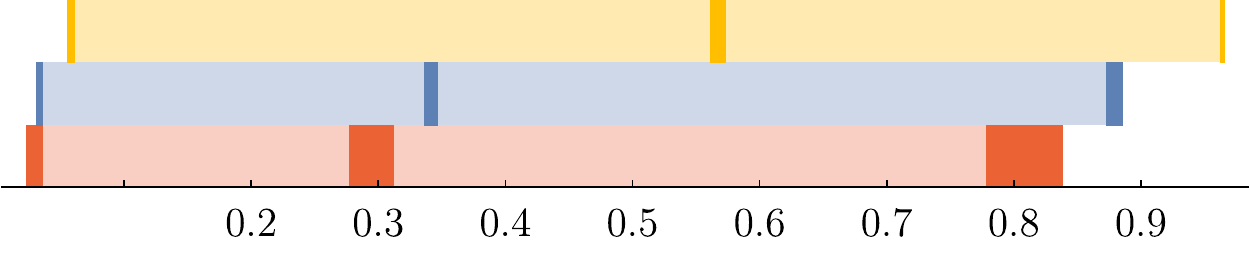}&$a_2$\\
\includegraphics[width=0.7\columnwidth]{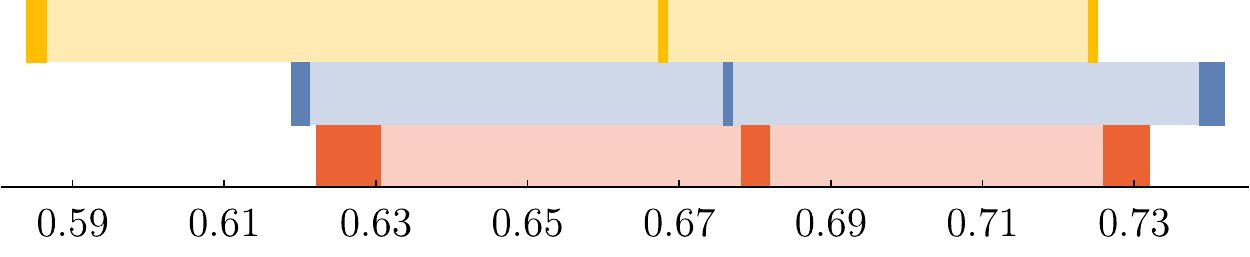}&$a_{\textrm{f}}$&&\\
\includegraphics[width=0.7\columnwidth]{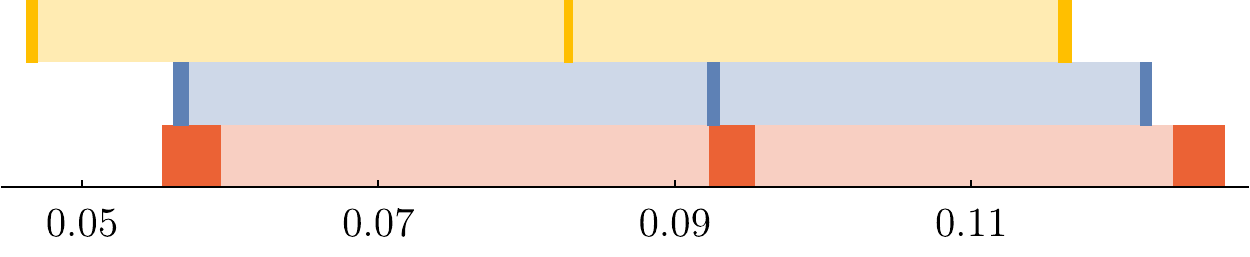}&$z$&\includegraphics[width=0.7\columnwidth]{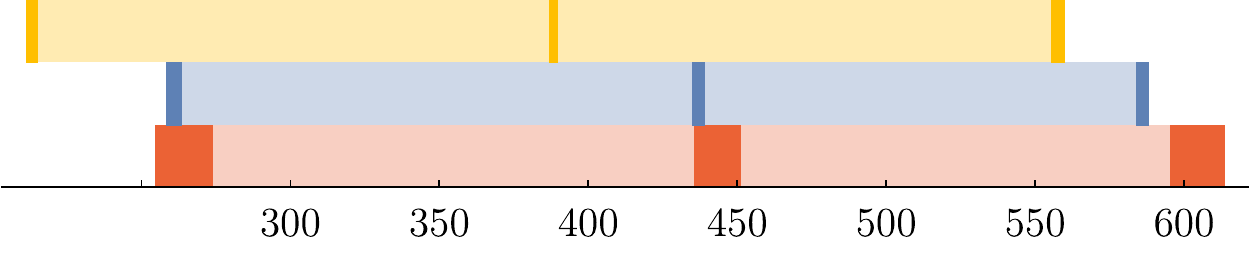}&$D_{\textrm{L}}/\textrm{Mpc}$\\
\includegraphics[width=0.7\columnwidth]{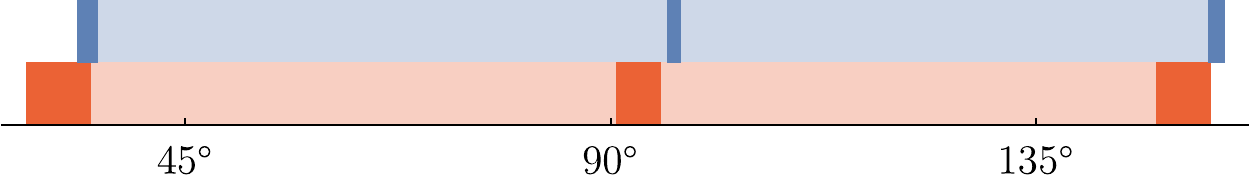}&$\theta_{LS_{1}}$&\includegraphics[width=0.7\columnwidth]{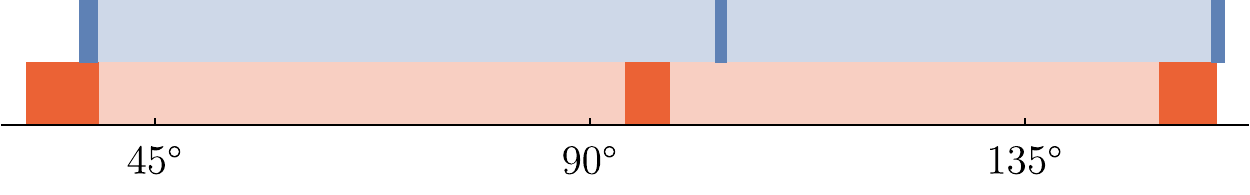}&$\theta_{LS_{2}}$\\
\includegraphics[width=0.7\columnwidth]{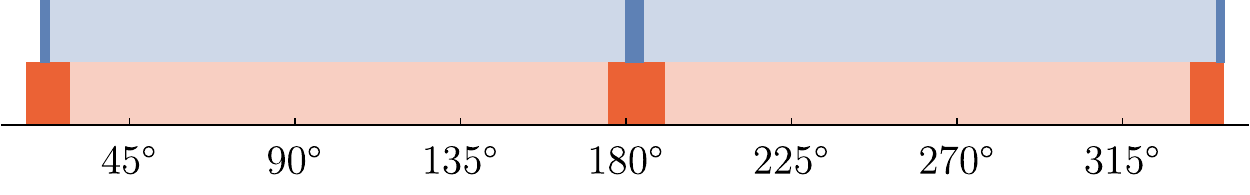}&$\phi_{12}$&&
\end{tabular}
\label{fig:tableModels}
\end{figure*}

The first question that we address is whether parameter estimates performed using the two precessing models (precessing IMRPhenom and precessing EOBNR) are compatible.
In particular, we wish to compare medians and 90\% credible intervals (the summary statistics used in Ref.\ \cite{GW150914-PARAMESTIM}) for the parameters tabulated in Table~I of Ref.~\cite{GW150914-PARAMESTIM}, as well as additional spin parameters.
The nominal values of the medians and $5\%$ and $95\%$ quantiles for the two models are listed in the ``EOBNR'' and ``IMRPhenom'' columns of Table~\ref{tab:parametersOld}.
However, it is unclear \textit{a priori} whether any differences are due to the models themselves, or to the imperfect sampling of the posteriors in Markov Chain Monte Carlo runs.
This is a concern especially for the precessing EOBNR results, since the slower speed of EOBNR waveform generation means that shorter chains are available for parameter estimation.
To gain trust in our comparisons, we characterize the Monte Carlo error of the medians and quantiles by a bootstrap analysis, as follows.

The Monte Carlo runs for the precessing IMRPhenom model produced an equal-weight posterior sampling consisting of 27,000 approximately independent samples, obtained by downsampling the original MCMC run by a factor equal to the largest autocorrelation length measured for the parameters of interest (those of Table~\ref{tab:parametersOld}).
We generate 1,000 Bayesian-bootstrap weighted resamplings \cite{rubin1981} of the equal-weight population,\footnote{For $n$ samples, this involves generating 1,000 realizations of weights according to the $(n - 1)$-variate Dirichlet distribution.} and for each we compute the weighted medians and quantiles. We characterize the Monte Carlo error of these summary statistics as the 90\% symmetric interquantile interval across the 1,000 realizations. For completeness, we apply the same analysis to the 45,000 samples of the nonprecessing EOBNR that were employed in Ref.\ \cite{GW150914-PARAMESTIM}.

The Monte Carlo runs for the precessing EOBNR model produced a sampling of 2,700 approximately independent samples, obtained by selecting every 11th sample in the original MCMC run. Again we generate 1,000 Bayesian-bootstrap resamplings, compute summary statistics on each, and measure their variation. However, to improve the representativeness of this analysis given the smaller number of samples in play, we use 9 additional equal-weight populations, obtained by selecting every $(11+i)$-th sample in the original MCMC run, for $i = 1, \ldots, 9$. For each of the 1,000 Bayesian-bootstrap resamplings, we first choose randomly among the 10 equal-weight populations.

Monte Carlo errors are expected to shrink as the inverse square root of the number of samples; this is indeed what we observe, with precessing EOBNR finite-sample errors $\sim (\mbox{27,000}/\mbox{2,700})^{1/2} \approx 3$ times larger than for precessing IMRPhenom. Table~\ref{tab:parametersOld} and Figure~\ref{fig:tableModels} present the results of this study for several key physical parameters of the source of GW150914.  We display with darker colors the finite-sample error estimates on the position of the medians and $5\%$ and $95\%$ quantiles. Lighter colors represent the 90\% credible intervals.

\paragraph*{Combined estimates.} To account for waveform-mismodeling errors in its fiducial parameter estimates, Ref.\ \cite{GW150914-PARAMESTIM} cited quantiles for combined posteriors obtained by averaging the posteriors for its two models (in Bayesian terms, this corresponds to assuming that the observed GW signal could have come from either model with equal prior probability).
We repeat the same procedure for the two precessing models, and we show the resulting estimates in the column ``Overall'' of Table \ref{tab:parametersOld}.
Quantiles are more uncertain for the precessing combination due to the larger finite-sampling error of precessing EOBNR. Nevertheless, 90\% credible intervals are slightly tighter than cited in Ref.\ \cite{GW150914-PARAMESTIM}.
In the Appendix, we provide a graphical representation of the combined estimates.

\begin{figure}
 \centering
 \includegraphics[width=0.78\columnwidth]{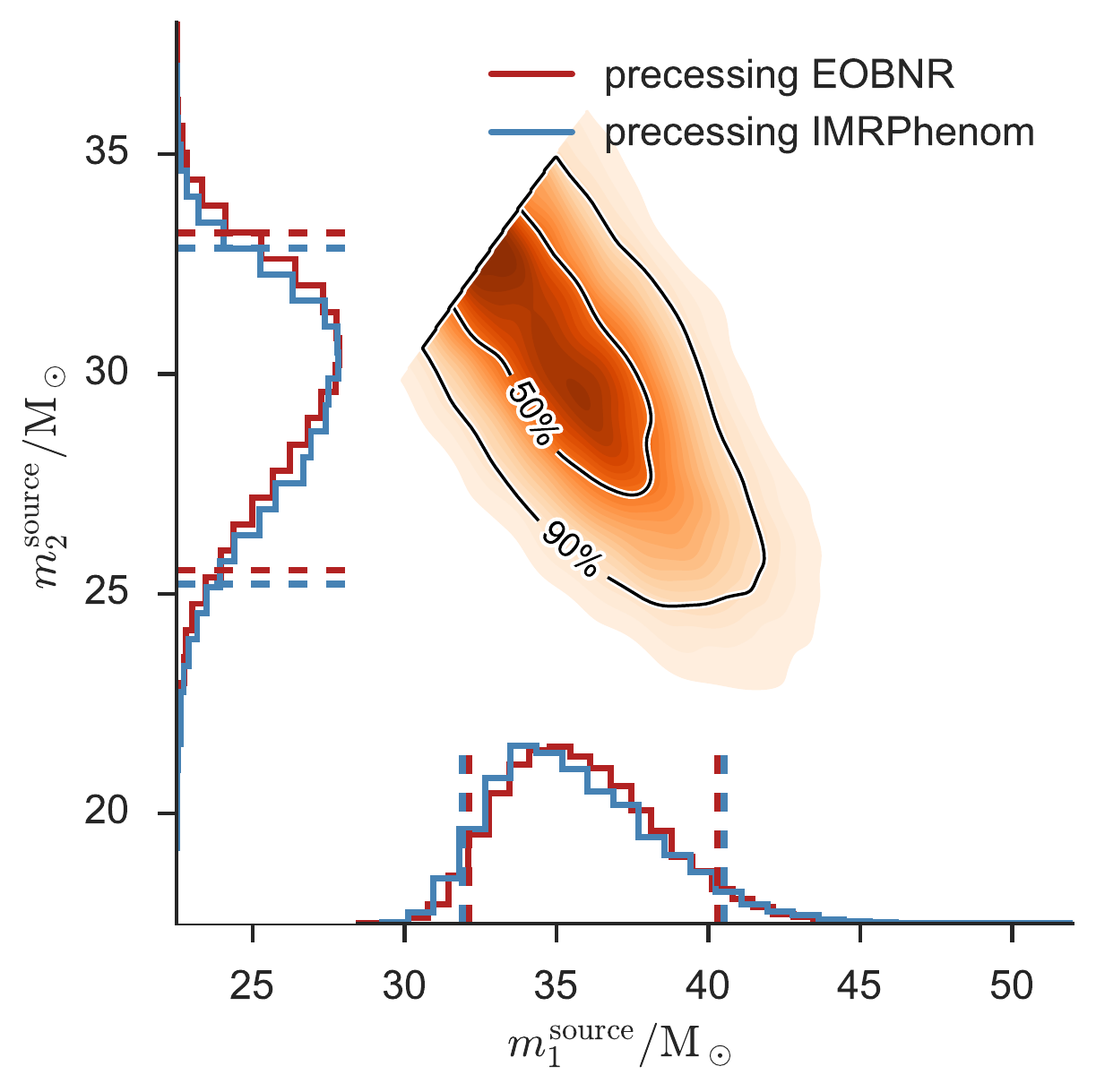}
 \caption{Posterior probability densities for the source-frame component
masses $m_1^{\mathrm{source}}$ and $m_2^{\mathrm{source}}$, where $m_2^{\mathrm{source}} \le m_1^{\mathrm{source}}$. We show one-dimensional histograms for precessing EOBNR (red) and precessing IMRPhenom (blue); the dashed vertical lines mark the $90\%$ credible intervals. The two-dimensional density plot shows $50\%$ and $90\%$ credible regions plotted over a color-coded posterior density function.
}
\label{fig:mass}
\end{figure}

\begin{figure}
 \centering
 \includegraphics[width=0.78\columnwidth]{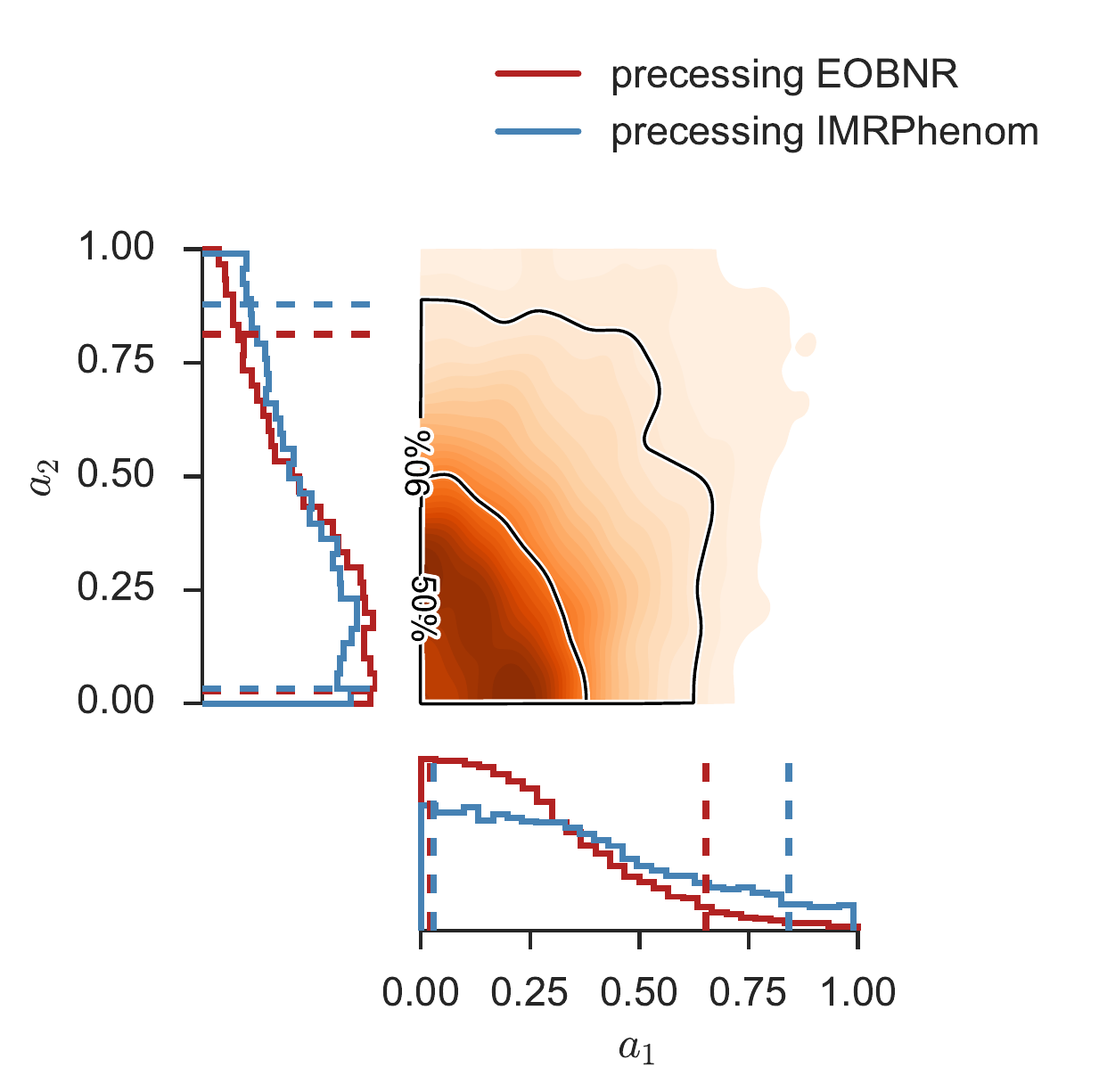}
 \caption{Posterior probability densities for the dimensionless spin magnitudes. (See Fig.\ \ref{fig:mass} for details.)
}
\label{fig:spin}
\end{figure}

\begin{figure}
 \centering
 \includegraphics[width=0.78\columnwidth]{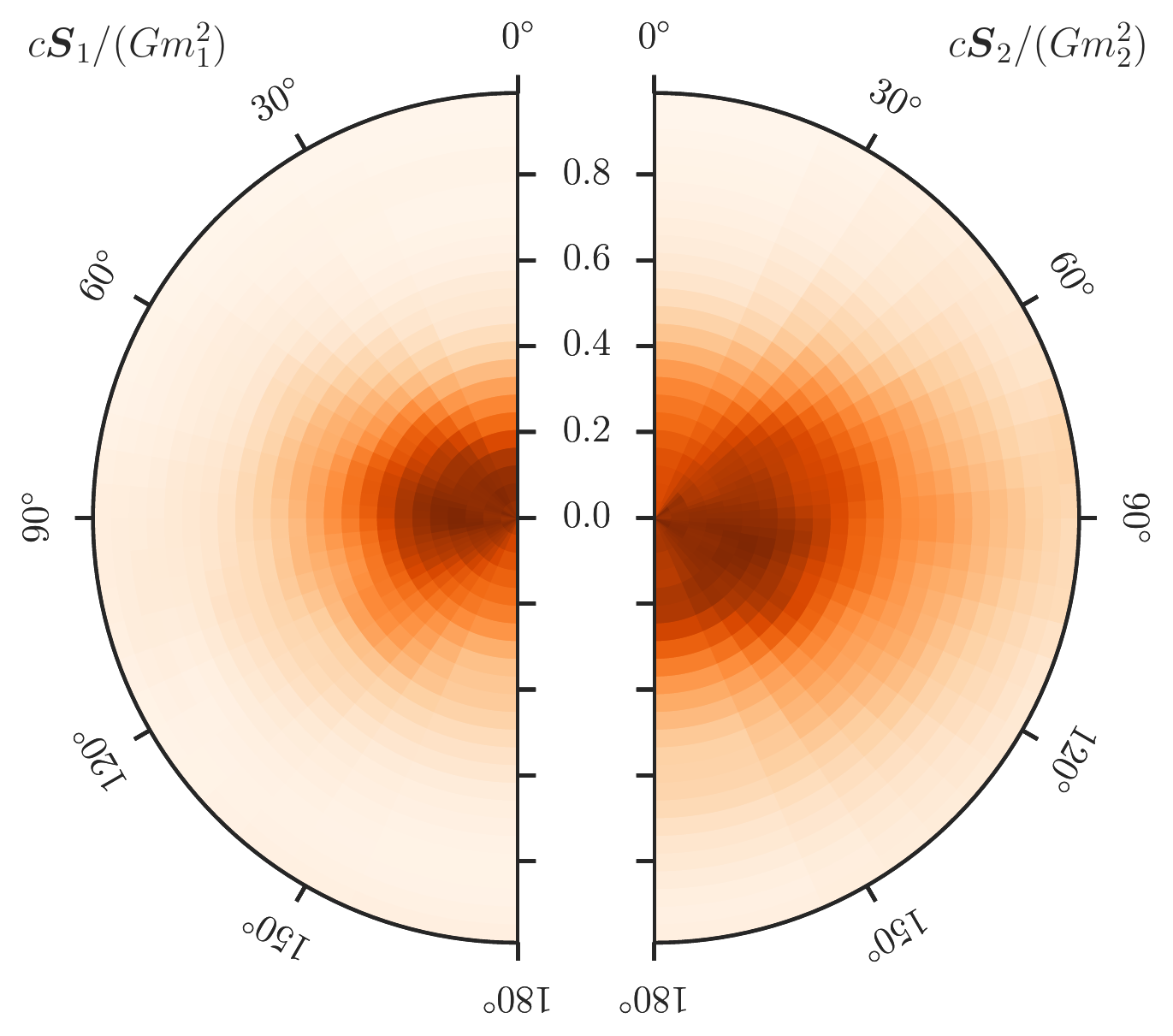}
 \caption{Posterior probability density of BH spin directions, plotted as
 in Fig.\ 5 of Ref.\ \cite{GW150914-PARAMESTIM}.
}
\label{fig:spin_disk}
\end{figure}

\begin{figure}
 \centering
 \includegraphics[width=0.78\columnwidth]{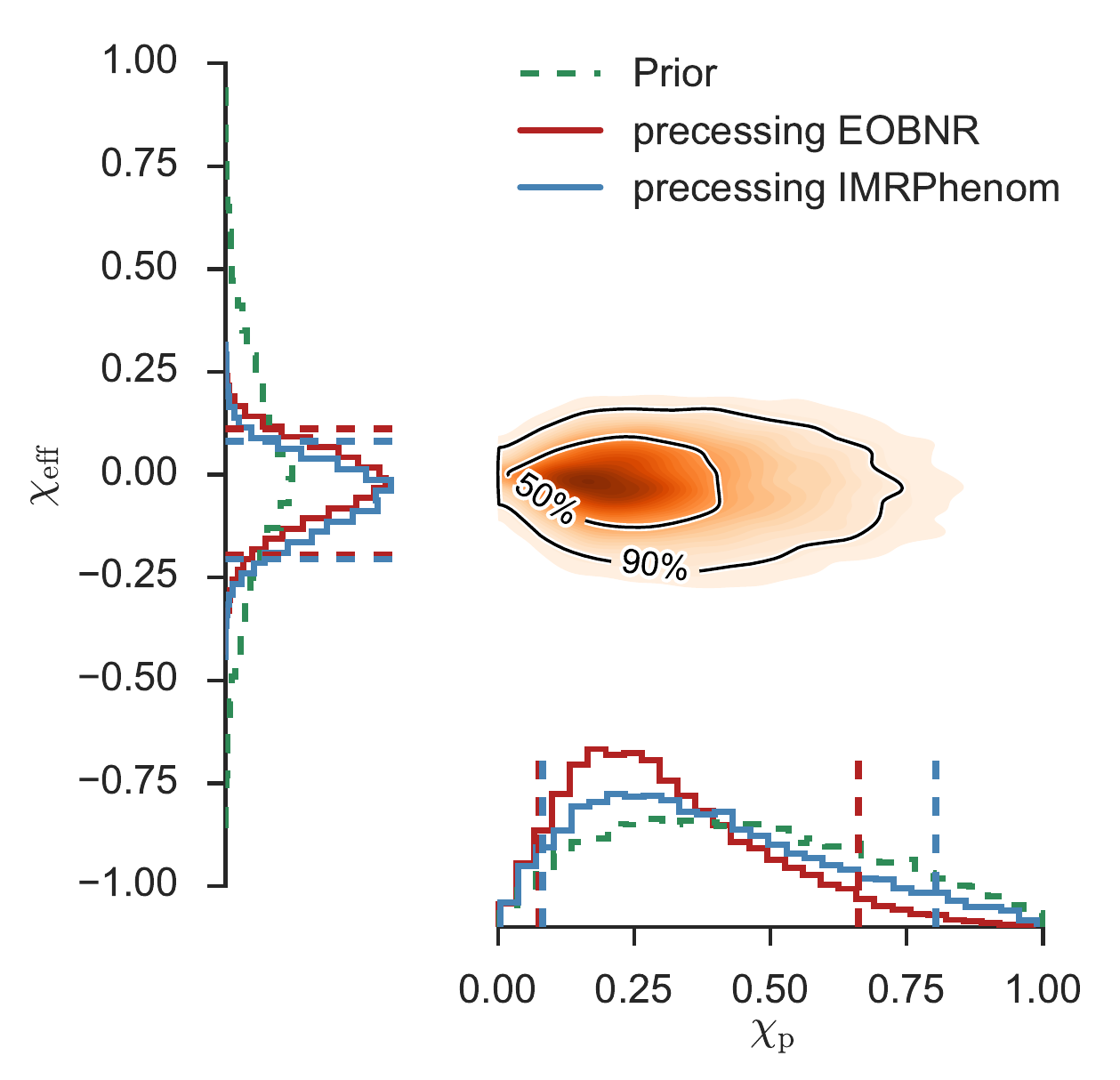}
 \caption{Posterior probability densities of the effective spin and perpendicular effective spin. (See Fig.\ \ref{fig:mass} for details.)
}
\label{fig:eff_spins}
\end{figure}

\begin{figure}
 \centering
 \includegraphics[width=0.78\columnwidth]{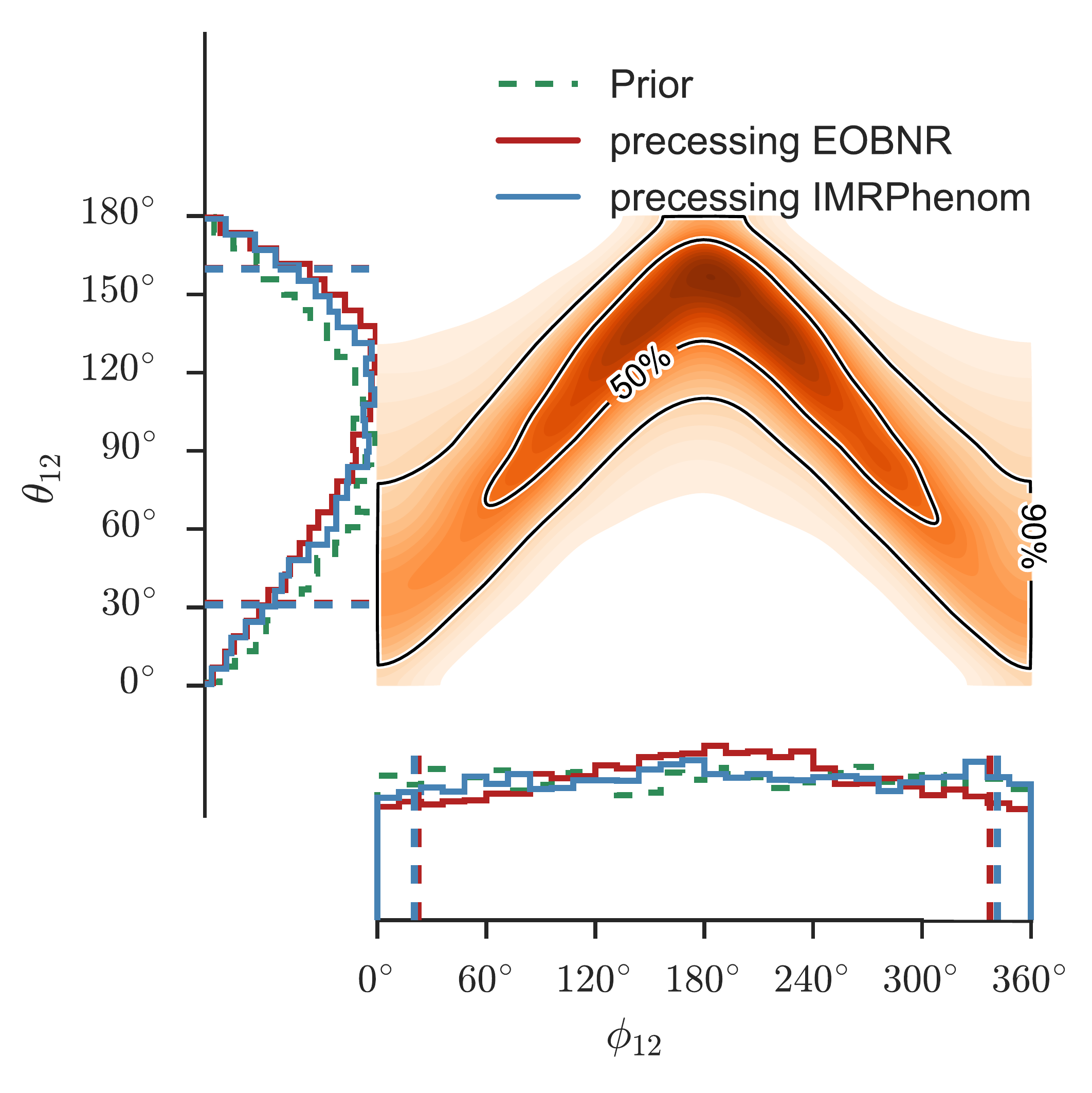}
 \caption{Posterior probability densities of the opening angle between the two spins, $\theta_{12}$, and the angle between the in-plane projections of the spins, $\phi_{12}$. (See Fig.\ \ref{fig:mass} for details.)
}
\label{fig:spinspin}
\end{figure}

\paragraph*{Posterior histograms: masses and spin magnitudes.} We now discuss in some detail the salient features of parameter posteriors. In Figs.\ \ref{fig:mass}--\ref{fig:spinspin}, we show the one-dimensional marginalized posteriors for selected pairs of parameters and 90\% credible intervals (the dashed lines), as obtained for the two precessing models, as well as the two-dimensional probability density plots for the precessing EOBNR model.
In Fig.\ \ref{fig:mass}, we show the posteriors for the \emph{source-frame} BH masses $m_{1,2}$: these are measured fairly well, with statistical uncertainties around 10\%.
In Fig.~\ref{fig:spin}, we show the posteriors for the dimensionless spin magnitudes $a_{1,2}$: the bound on $a_1$ is about 20\% more stringent for precessing EOBNR. This is true even if we account for the larger finite-sampling uncertainty in the precessing EOBNR quantiles (see Table \ref{fig:tableModels}).
The final spin presented in Table~\ref{tab:parametersOld} and Figure~\ref{fig:tableModels} was obtained including the contribution from the in-plane spin components to the final spin~\cite{spinfit-T1600168}; previous publications~\cite{GW150914-DETECTION, GW150914-PARAMESTIM} just use the contribution from the aligned components of the spins, which remains sufficient for the final mass computation. Just using the aligned components does not change the precessing EOBNR result, but gives a precessing IMRPhenom result of $0.66^{+0.04}_{-0.06}$.

\paragraph*{Posterior histograms: spin directions.}
Figure \ref{fig:spin_disk} reproduces the disk plot of Ref.\ \cite{GW150914-PARAMESTIM} for precessing EOBNR. In this plot, the three-dimensional histograms of the dimensionless spin vectors $\vS_{1,2}/m_{1,2}^2$ are projected onto a plane perpendicular to the orbital plane; the bins are designed so that each contains the same prior probability mass (i.e., histogramming the prior would result in a uniform shading).
It is apparent that the data disfavor large spins aligned or antialigned with the orbital angular momentum, consistently with precessing IMRPhenom results.
Because precessing EOBNR favors smaller values of the dimensionless spin magnitudes, the plot is darker toward its center than its counterpart in Ref.\ \cite{GW150914-PARAMESTIM}.
In agreement with that paper, our analysis does not support strong statements on the orientation of the BH spins with respect to the orbital angular momentum.  
The spin opening angles (the \emph{tilts}), defined by $\cos(\theta_{LS_{1,2}}) = (\vS_{1,2}\cdot \vLhat_{\text{N}})/|\vS_{1,2}|$, are distributed broadly.
However, the KS test described at the end of Sec.\ \ref{S:Description} does indicate some deviation between priors and posteriors, with $p$-values much smaller than 0.05 for $\cos(\theta_{LS_{1}})$ and $\cos(\theta_{LS_{2}})$.

\paragraph*{Posterior histograms: effective spin parameters.}
In Fig.\ \ref{fig:eff_spins}, we show the posteriors of the effective spin combinations $\chi_{\textrm{eff}}$~\cite{Damour:2001tu,Racine:2008qv,Ajith:2009bn,Santamaria:2010} and $\chi_{\textrm{p}}$~\cite{Schmidt:2014iyl} defined by 

\begin{eqnarray}
\label{effspins}
\chi_\mathrm{eff} &=&  \frac{c}{G}\left(\frac{\boldsymbol{S}_1}{m_1} +
\frac{\boldsymbol{S}_2}{m_2} \right) \cdot \frac{\boldsymbol{\hat L_{\text{N}}}}{M}\,, \\
\chi_\mathrm{p} &=& \frac{c}{B_1 G m_1^2}{{\max} (B_1 S_{1 \perp}, B_2 S_{2 \perp})}\,,
\end{eqnarray}
where $\boldsymbol{S}_{i\perp}$ is 
the component of the spin perpendicular to the orbital angular momentum $\boldsymbol{L_{\text{N}}}$, $M$ is the total observed mass, 
$B_1 = 2 + 3q/2$ and $B_2 = 2+ 3/(2q)$, and $i = \{1,2\}$. 

While $\chi_{\textrm{eff}}$ combines the projections of the BH spins onto the orbital angular momentum, $\chi_{\textrm{p}}$ depends on their in-plane components, and thus relates to precessional effects.
Both models have credible intervals for $\chi_{\textrm{eff}}$ that contain the value 0, and deviate from the prior significantly. The data provides little information about precession, but show a slightly stronger preference for lower values of $\chi_\textrm{p}$ than expressed by our priors; the deviation is more pronounced for precessing EOBNR.
The 90\% credible intervals contain the value 0, and extend up to about 0.7 and 0.8 for precessing EOBNR and precessing IMRPhenom, respectively. Thus, precessing EOBNR provides a tighter upper bound.

\paragraph*{Posterior histograms: other spin angles.}
To explore other possible differences between the two precessing models, we now consider spin parameters that were not reported in Ref.~\cite{GW150914-PARAMESTIM}. In particular, we compute posteriors for $\theta_{12}$, the opening angle between the spin vectors, and $\phi_{12}$, the opening angle between the in-plane projections of the spins. The prior on $\cos{\theta_{12}}$ is uniform in $[-1,1]$, while the prior on $\phi_{12}$ is uniform in $[0,2\pi]$. We show these posteriors in Fig. \ref{fig:spinspin}. 
The $\theta_{12}$ posteriors deviate appreciably from the prior, and are rather similar. By contrast, comparing the opening angle between spin projections onto orbital plane, $\phi_{12}$, we find that the precessing EOBNR posterior deviates significantly from the prior (with KS $p$-value $\in [0.0077,0.075]$), while the precessing IMRPhenom posterior  does not (with KS $p$-value $\in [0.30,0.60]$). This is as it should be, since in precessing IMRPhenom binaries with identical projection of the total spin on the orbital plane have identical waveforms. Although the KS $p$-values suggest that the data provide information about $\theta_{12}$ and $\phi_{12}$ beyond the prior, we note that the 90\% confidence intervals for both of these parameters cover approximately 90\% of their valid ranges, and are indistinguishable for each waveform model.

\paragraph*{Spin evolution}
All the source parameters discussed above are measured at a reference frequency of 20~Hz. Exploiting the capability of precessing EOBNR of evolving the BH spin vectors in the time domain, we may address the question of estimating values for the spin parameters at the time of the merger. To do so, we randomly sample 1,000 distinct configurations from the precessing EOBNR posteriors, and we evolve them to the maximum EOB orbital frequency (a proxy for the merger in the model). We then produce histograms of the evolved values of $\chi_{\textrm{eff}}$ and $\chi_{\textrm{p}}$. We find little if any change between 20~Hz and the merger. Indeed, a KS test suggests that the original and evolved distributions are very consistent, with $p$-values close to 1.

\paragraph*{Comparison with numerical relativity}
The precessing EOBNR waveform model has been tested against \ac{NR} waveforms using simulations from the SXS catalog~\cite{Pan:2013rra,BabaketalInPrep,Mroue:2013xna}. We can provide a targeted cross-check on the accuracy of precessing EOBNR near GW150914 by performing parameter estimation runs on mock \ac{NR} signals injected into LIGO data. This test is complementary to an ongoing study of the same nature that however does not employ the precessing EOBNR model used in this paper. We use a new \ac{LAL} infrastructure~\cite{NRinjSchmidtHarryInPrep,NRSplineSchmidtGalleyInPrep} to inject spline-interpolated and tapered \ac{NR} waveforms into detector data; spins are defined with respect to the orbital angular momentum at a reference frequency of $20$ Hz. All higher harmonics of the GW signal are included up to the $l=8$ multipole. 
At the inclinations used in this study the impact of modes with $l>2$ is small, but merits further study, a detailed analysis will be presented in a forthcoming paper.  
We restrict this investigation to a \ac{NR} waveform which was computed by the SXS collaboration using the SpEC~\cite{SpECwebsite} code and is available in the public waveform catalogue~\cite{SXSCatalog} as \texttt{SXS:BBH:0308}. The intrinsic parameters of the \ac{NR} waveform $q=0.81$, $a_1=0.34$, and $a_2=0.67$ are  consistent with the results obtained in Ref.~\cite{GW150914-PARAMESTIM} and this waveform agrees well with the detector data.

\begin{figure*}
  \centering
    \includegraphics[width=0.7\linewidth]{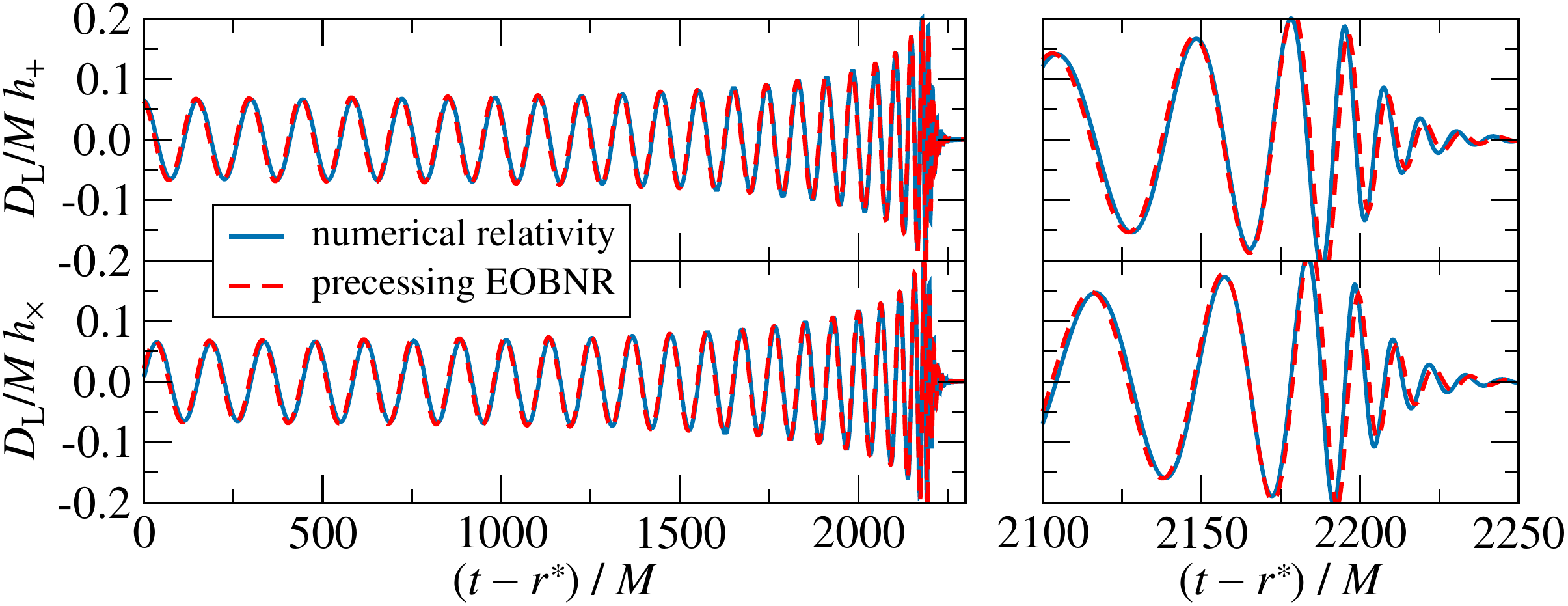}
  \caption{A visual comparison of the precessing EOBNR model with NR GW polarizations computed by the SXS collaboration at (approximately) the GW150914 \ac{MaP} parameters. The intrinsic parameters of the \ac{NR} waveform are $q=0.81$, $a_1=0.34$, $a_2=0.67$. The inclination is $\iota = 2.856$~rad. The alignment of the precessing EOBNR waveform is obtained from the sky- and polarization-averaged overlap with the \ac{NR} waveform.}
  \label{fig:HpHc}
\end{figure*}

We can freely choose the angle between the line of sight and the angular momentum of the binary for mock \ac{NR} signals. Since there is some uncertainty in the binary's inclination, we perform one run near \ac{MaP} inclination, $\iota = 2.856$~rad ($163.6 ^{\circ}$), and a second one at the upper bound of the 90\% credible interval of the marginal \ac{PDF} of the inclination, $\iota = 1.2$ rad ($68.8 ^{\circ}$). In Fig.~\ref{fig:HpHc} we show the two GW polarizations for the \ac{NR} waveform and the precessing EOBNR model. The spin magnitudes and the mass ratio were fixed to the \ac{NR} values. The directions of spins are defined to be the same at initial time: tilt angles are $18.8^{\circ}, \; 149.4^{\circ}$, and the azimuthal angles, defined with respect to the initial separation vector, are $30.9^{\circ},\;  38.7^{\circ}$ for the primary and secondary BHs, respectively. 
To quantify the agreement between those waveforms we compute overlaps averaged over the GW polarization and source sky location, which 
takes into account the uncertainty in those parameters. The polarization-sky-averaged overlap for \ac{MaP} inclination  is 0.997, and for $\iota = 1.2$ rad ($68.8 ^{\circ}$)  overlap is 0.993.

\begin{figure*}[!]
  \centering
   \begin{tabular}{cc}
    \includegraphics[width=0.7\columnwidth]{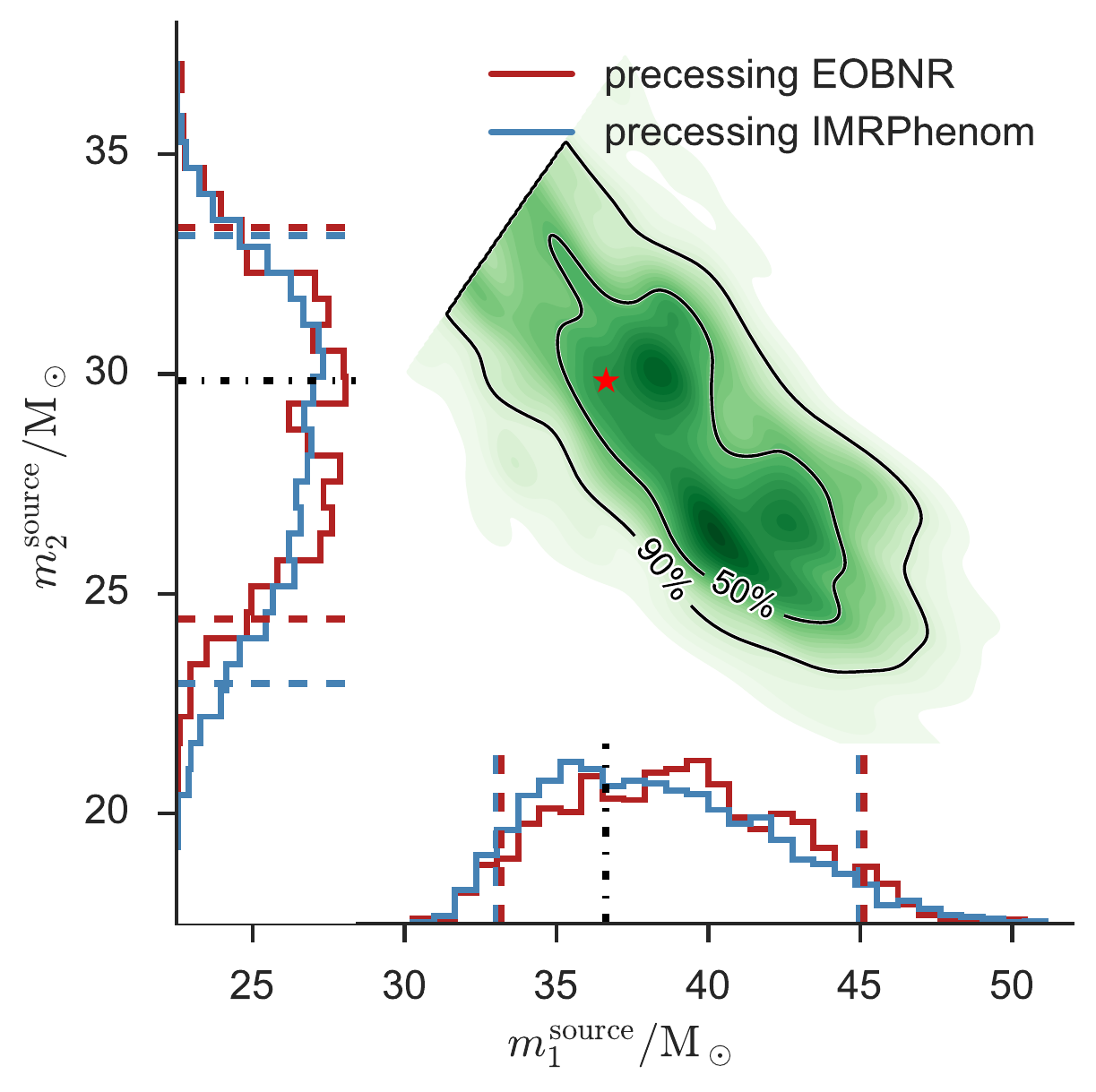} \hspace{1truecm}
    \includegraphics[width=0.7\columnwidth]{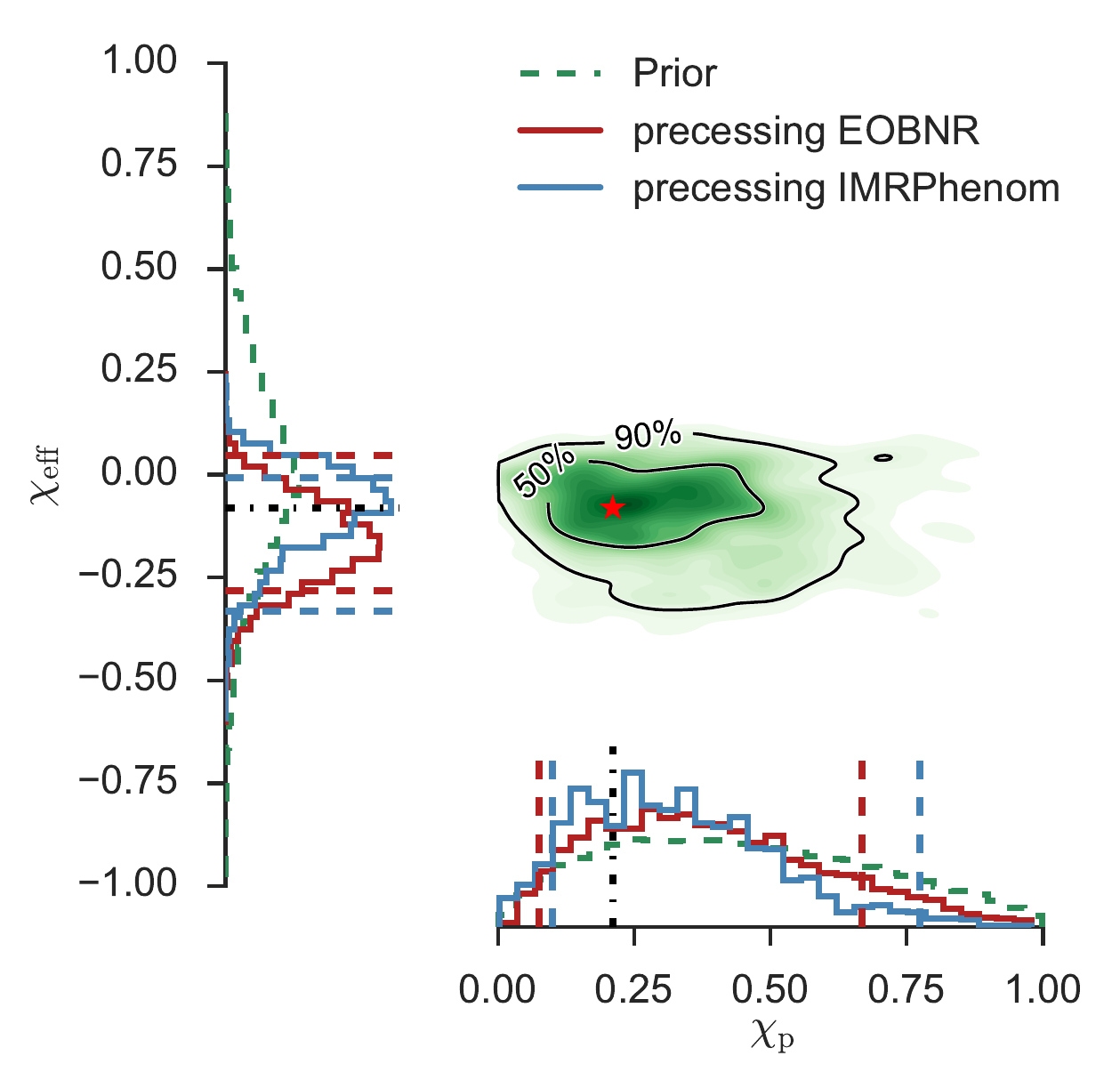}
   \end{tabular} 
  \caption{Posterior probability densities for the source-frame component masses $m_1^{\mathrm{source}}$ and $m_2^{\mathrm{source}}$, where $m_2^{\mathrm{source}} \le m_1^{\mathrm{source}}$, (left) and effective aligned $\chi_\mathrm{eff}$ and effective precessing spins $\chi_\mathrm{p}$ (right) for an event-like \ac{NR} mock signal close to \ac{MaP} parameters. In the $1$-dimensional marginalized distributions we show the precessing EOBNR (red) and precessing IMRPhenom (blue) probability densities with dashed vertical lines marking $90\%$ credible intervals. The $2$-dimensional plot shows the contours of the $50\%$ and $90\%$ credible regions of the precessing EOBNR over a color-coded posterior density function. The true parameter values are indicated by a red asterisk or black dot-dashed line.}
  \label{fig:figs_NR_SO_prec_MaP}
\end{figure*}

We show results for the run with \ac{MaP} inclination for the source-frame component masses and effective spins in the left and right panels of Fig.~\ref{fig:figs_NR_SO_prec_MaP}. The precessing EOBNR and precessing IMRPhenom model show good agreement in the masses and effective precession spin $\chi_\mathrm{p}$. The posterior \ac{PDF}s obtained for the effective aligned spin $\chi_\mathrm{eff}$ are slightly offset. All injected values are found within the $90\%$ credible regions.
Results for the inclination chosen at the upper bound of the 90\% credible interval of the marginal \ac{PDF} of the inclination are qualitatively similar to the \ac{MaP} results, except for the \ac{PDF} of the effective precession spin which peaks around $\chi_\mathrm{p} \sim 0.4$, noticeably above the injected value, but still well inside the $90\%$ credible interval.

\section{Conclusions} 
\label{S:Conclusion}

We presented an updated analysis of \TheEvent{} with mass estimates of $35^{+5}_{-3}$\ \Msun\ and $30^{+3}_{-4}$\ \Msun, and we refined parameter estimates using a generalized, fully precessing waveform model which depends on the full 15 independent parameters of a coalescing binary in circular orbit. We find this analysis to be broadly consistent with the results in Ref.~\cite{GW150914-PARAMESTIM}.
By using the difference between two precessing waveform models as a proxy for systematic errors due to waveform uncertainty, we can compute a more accurate systematic error than what was possible in Ref.~\cite{GW150914-PARAMESTIM}. By looking at differences in 5\% and 95\% quantiles between different waveform models in Fig.~\ref{fig:tableModels}, one can observe, on average, more consistent values when the two precessing models are compared. In addition, this analysis provides an estimate of the systematic error on precessing spin parameters such as the effective precessing spin $\chi_\mathrm{p}$ and the tilt angles ($\mbox{\boldmath${\hat{S}}$}_{1,2}\cdot \mbox{\boldmath${\hat{L}}$}_{\text{N}}$), which was not available in Ref.~\cite{GW150914-PARAMESTIM}. We have also carefully investigated uncertainties due to the finite numbers of samples used to recreate continuous posterior density functions, and we quantified their effects on quoted credible intervals. As in Ref.~\cite{GW150914-PARAMESTIM}, the statistical error due to finite signal-to-noise ratio dominates the parameter-estimation error.

While we do recover a tighter limit on the spin magnitude of the most massive member of the binary that created \TheEvent~($<0.65$ at 90\% probability), the recovery of the spin parameters (magnitude and tilt angles) is too broad to hint at whether the black hole binary was formed via stellar binary interactions or dynamical capture~\cite{GW150914-ASTRO}.
This analysis on the first direct detection by LIGO will be applied to future detections~\cite{GW150914-RATES}, with the aim of getting the most accurate and most precise parameter estimate possible. In particular, binaries that have larger mass asymmetry, that are observed for a longer time, and that are more edge-on than GW150914 will display stronger spin-precession effects.

\acknowledgments 
The authors gratefully acknowledge the support of the United States
National Science Foundation (NSF) for the construction and operation of the
LIGO Laboratory and Advanced LIGO as well as the Science and Technology Facilities Council (STFC) of the
United Kingdom, the Max-Planck-Society (MPS), and the State of
Niedersachsen/Germany for support of the construction of Advanced LIGO 
and construction and operation of the GEO\,600 detector. 
Additional support for Advanced LIGO was provided by the Australian Research Council.
The authors gratefully acknowledge the Italian Istituto Nazionale di Fisica Nucleare (INFN),  
the French Centre National de la Recherche Scientifique (CNRS) and
the Foundation for Fundamental Research on Matter supported by the Netherlands Organisation for Scientific Research, 
for the construction and operation of the Virgo detector
and the creation and support  of the EGO consortium. 
The authors also gratefully acknowledge research support from these agencies as well as by 
the Council of Scientific and Industrial Research of India, 
Department of Science and Technology, India,
Science \& Engineering Research Board (SERB), India,
Ministry of Human Resource Development, India,
the Spanish Ministerio de Econom\'ia y Competitividad,
the Conselleria d'Economia i Competitivitat and Conselleria d'Educaci\'o, Cultura i Universitats of the Govern de les Illes Balears,
the National Science Centre of Poland,
the European Commission,
the Royal Society, 
the Scottish Funding Council, 
the Scottish Universities Physics Alliance, 
the Hungarian Scientific Research Fund (OTKA),
the Lyon Institute of Origins (LIO),
the National Research Foundation of Korea,
Industry Canada and the Province of Ontario through the Ministry of Economic Development and Innovation, 
the Natural Science and Engineering Research Council Canada,
Canadian Institute for Advanced Research,
the Brazilian Ministry of Science, Technology, and Innovation,
Russian Foundation for Basic Research,
the Leverhulme Trust, 
the Research Corporation, 
Ministry of Science and Technology (MOST), Taiwan
and
the Kavli Foundation.
The authors gratefully acknowledge the support of the NSF, STFC, MPS, INFN, CNRS and the
State of Niedersachsen/Germany for provision of computational resources.

\appendix*

\begin{figure*}
\caption{Comparison of parameter estimates obtained by combining the nonprecessing-EOBNR and precessing-IMRPhenom models (as in Ref.\ \cite{GW150914-PARAMESTIM}; light purple bars at the top) and by combining the precessing-EOBNR and precessing-IMRPhenom models (light green bars at the bottom). We show 90\% credible intervals for selected GW150914 source parameters. The darker intervals represent uncertainty estimates for the 5\%, 50\% and 95\% quantiles (from left to right), as estimated by the Bayesian bootstrap.}
\begin{tabular}{rlrl}
\includegraphics[width=0.6\columnwidth]{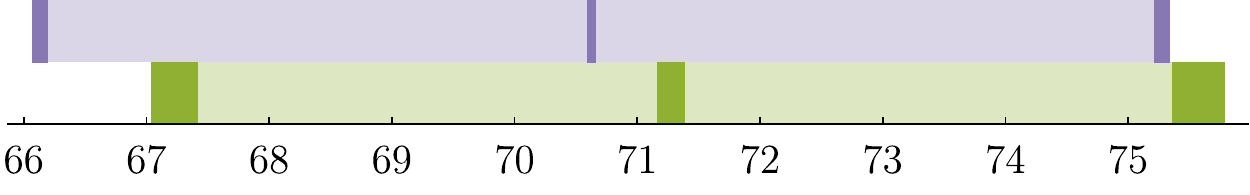}&$M/\textrm{M}_{\odot}$&\includegraphics[width=0.7\columnwidth]{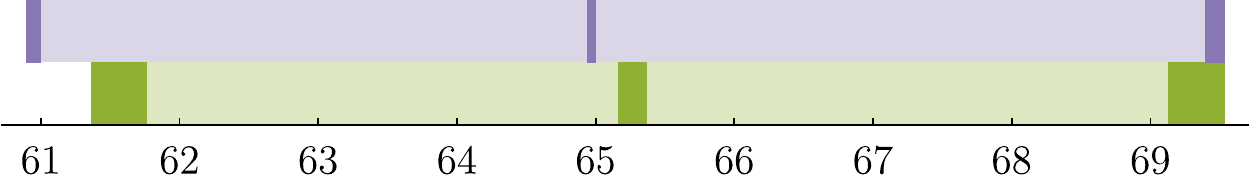}&$M^{\textrm{source}}/\textrm{M}_{\odot}$\\
\includegraphics[width=0.7\columnwidth]{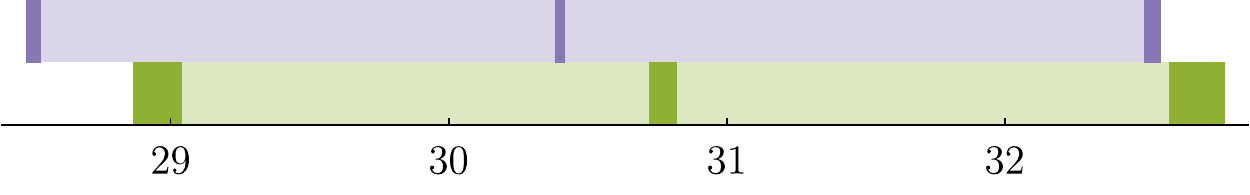}&$\mathcal{M}/\textrm{M}_{\odot}$&\includegraphics[width=0.7\columnwidth]{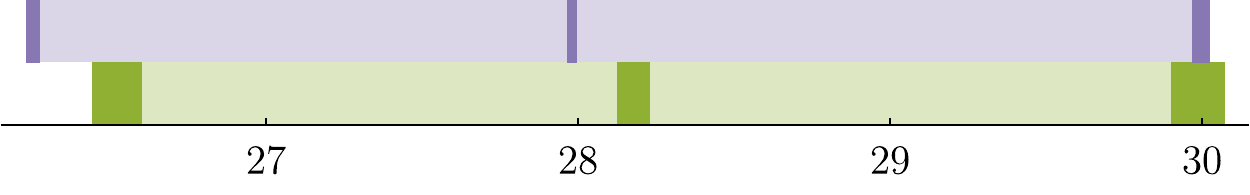}&$\mathcal{M}^{\textrm{source}}/\textrm{M}_{\odot}$\\
\includegraphics[width=0.7\columnwidth]{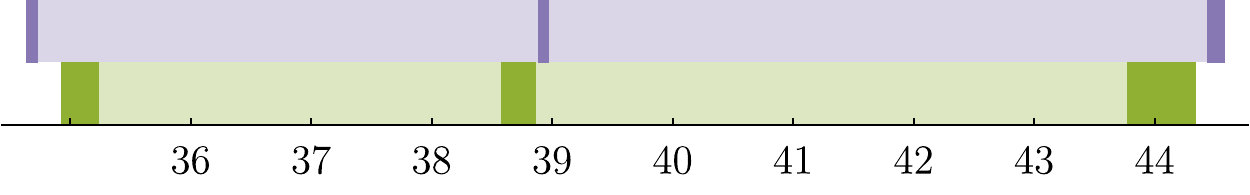}&$m_1/\textrm{M}_{\odot}$&\includegraphics[width=0.7\columnwidth]{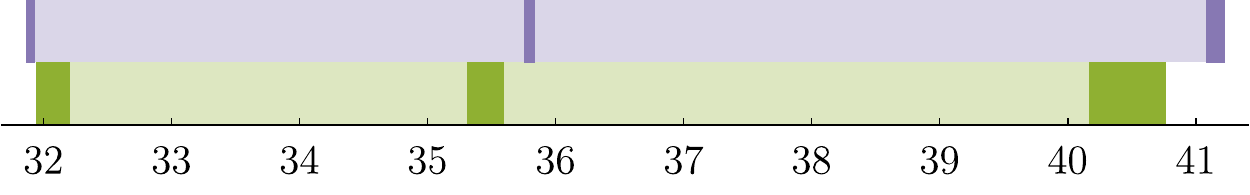}&$m_1^{\textrm{source}}/\textrm{M}_{\odot}$\\
\includegraphics[width=0.7\columnwidth]{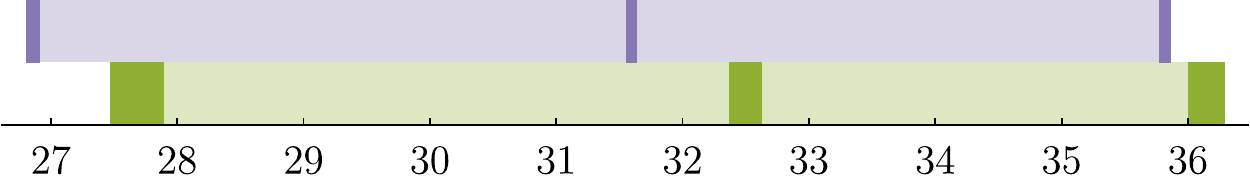}&$m_2/\textrm{M}_{\odot}$&\includegraphics[width=0.7\columnwidth]{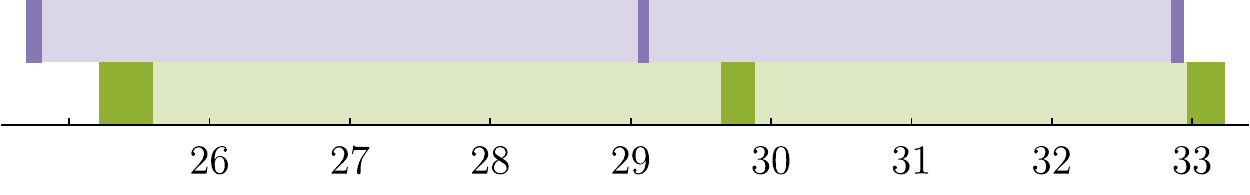}&$m_2^{\textrm{source}}/\textrm{M}_{\odot}$\\
\includegraphics[width=0.7\columnwidth]{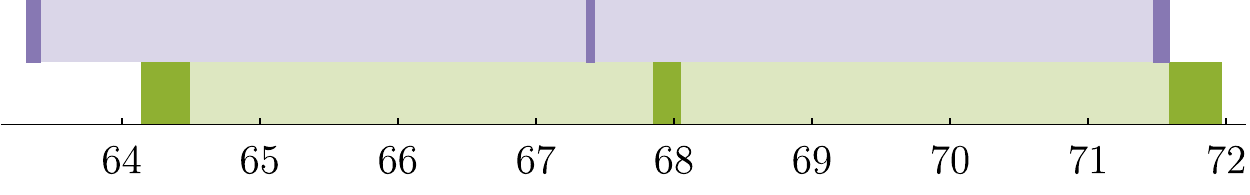}&$M_{\textrm{f}}/\textrm{M}_{\odot}$&
\includegraphics[width=0.7\columnwidth]{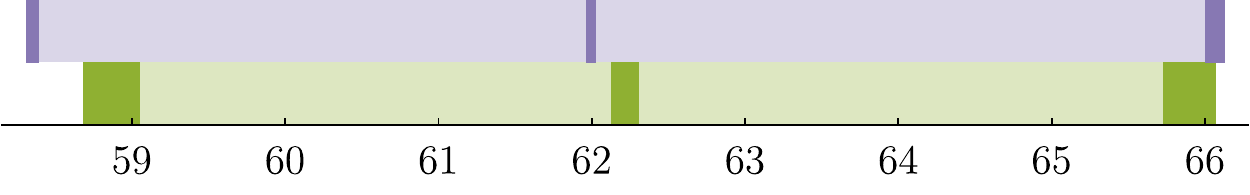}&$M_{\textrm{f}}^{\textrm{source}}/\textrm{M}_{\odot}$\\
\includegraphics[width=0.7\columnwidth]{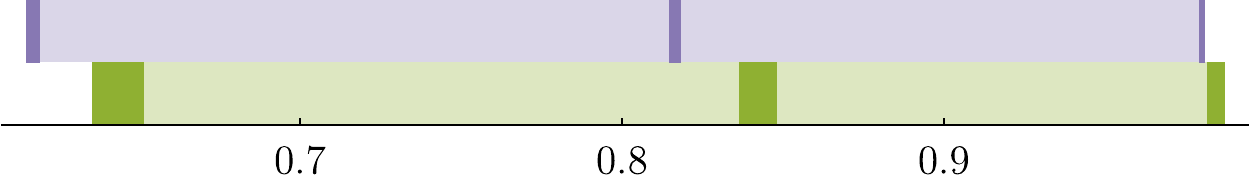} &$q$&&\\
\includegraphics[width=0.7\columnwidth]{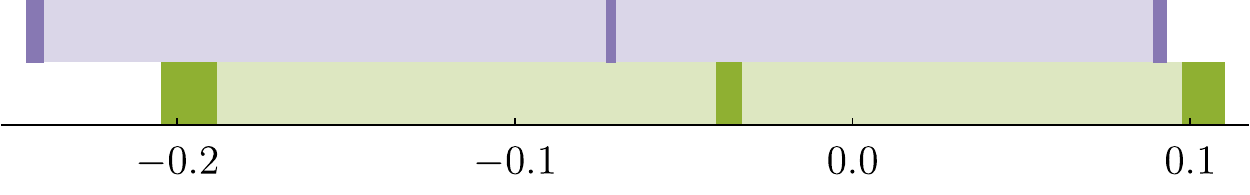}&$\chi_{\textrm{eff}}$&\includegraphics[width=0.7\columnwidth]{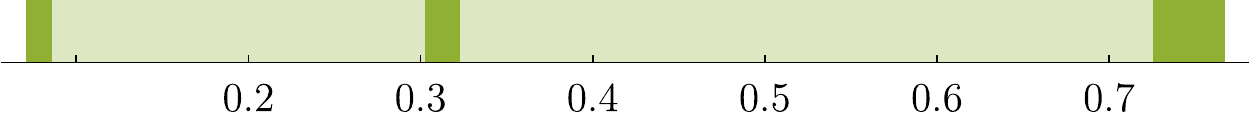}&$\chi_{\textrm{p}}$\\
\includegraphics[width=0.7\columnwidth]{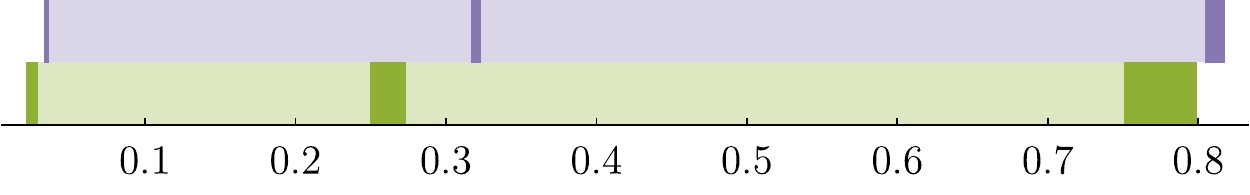}&$a_1$&\includegraphics[width=0.7\columnwidth]{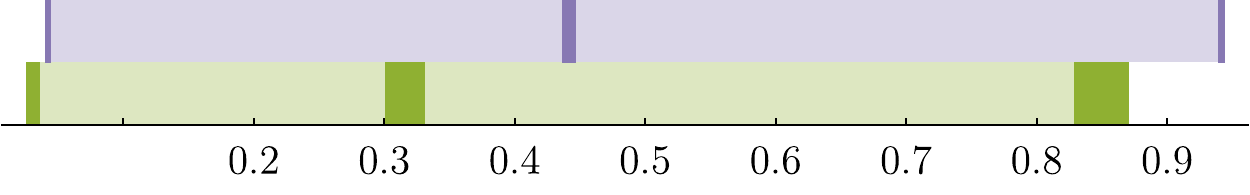}&$a_2$\\
\includegraphics[width=0.7\columnwidth]{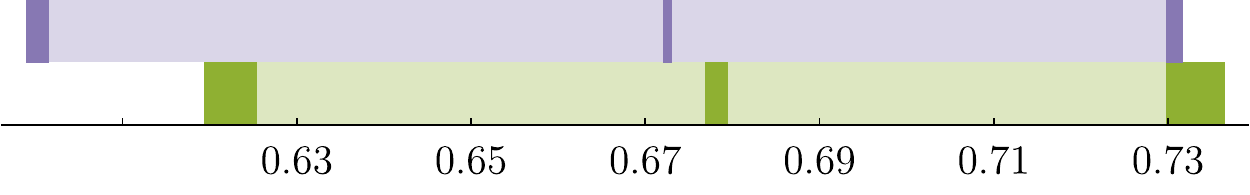}&$a_{\textrm{f}}$&&\\
\includegraphics[width=0.7\columnwidth]{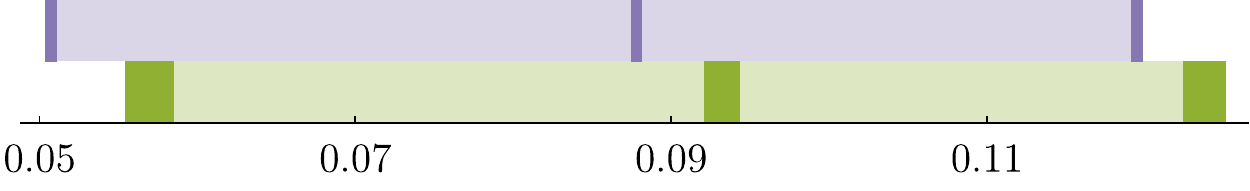}&$z$&\includegraphics[width=0.7\columnwidth]{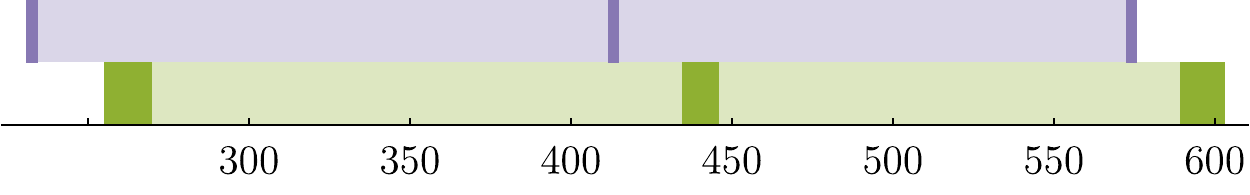}&$D_{\textrm{L}}/\textrm{Mpc}$\\
\includegraphics[width=0.7\columnwidth]{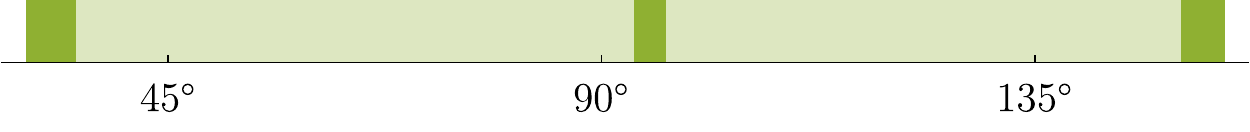}&$\theta_{LS_{1}}$&\includegraphics[width=0.7\columnwidth]{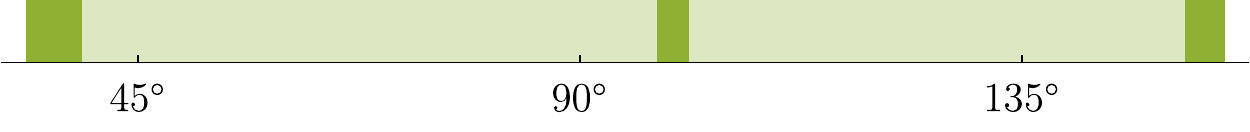}&$\theta_{LS_{2}}$\\
\includegraphics[width=0.7\columnwidth]{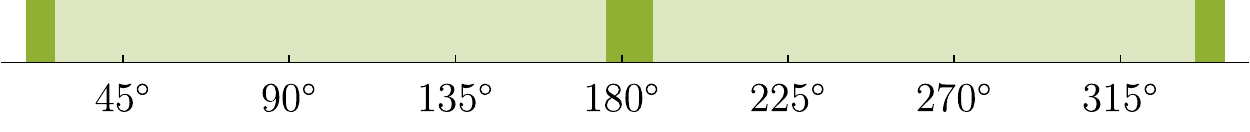}&$\phi_{12}$&&
\end{tabular}
\label{fig:tableCombined}
\end{figure*}

\section{Credible intervals for the combined posteriors}
To compare directly with the results of Ref.~\cite{GW150914-PARAMESTIM}, Table\,\ref{fig:tableCombined} presents the 90\% credible intervals obtained with combined nonprecessing-EOBNR and precessing-IMRPhenom models, and with combined precessing-EOBNR precessing-IMRPhenom models. As in Table\,\ref{fig:tableModels}, the darker bands visualize uncertainties due the finite number of samples, as estimated with the Bayesian bootstrap.
%

\FloatBarrier
\def\url{} 
\bibliography{references,refs,GW150914_refs}

\begin{thebibliography}{59}%
\makeatletter
\providecommand \@ifxundefined [1]{%
 \@ifx{#1\undefined}
}%
\providecommand \@ifnum [1]{%
 \ifnum #1\expandafter \@firstoftwo
 \else \expandafter \@secondoftwo
 \fi
}%
\providecommand \@ifx [1]{%
 \ifx #1\expandafter \@firstoftwo
 \else \expandafter \@secondoftwo
 \fi
}%
\providecommand \natexlab [1]{#1}%
\providecommand \enquote  [1]{``#1''}%
\providecommand \bibnamefont  [1]{#1}%
\providecommand \bibfnamefont [1]{#1}%
\providecommand \citenamefont [1]{#1}%
\providecommand \href@noop [0]{\@secondoftwo}%
\providecommand \href [0]{\begingroup \@sanitize@url \@href}%
\providecommand \@href[1]{\@@startlink{#1}\@@href}%
\providecommand \@@href[1]{\endgroup#1\@@endlink}%
\providecommand \@sanitize@url [0]{\catcode `\\12\catcode `\$12\catcode
  `\&12\catcode `\#12\catcode `\^12\catcode `\_12\catcode `\%12\relax}%
\providecommand \@@startlink[1]{}%
\providecommand \@@endlink[0]{}%
\providecommand \url  [0]{\begingroup\@sanitize@url \@url }%
\providecommand \@url [1]{\endgroup\@href {#1}{\urlprefix }}%
\providecommand \urlprefix  [0]{URL }%
\providecommand \Eprint [0]{\href }%
\providecommand \doibase [0]{http://dx.doi.org/}%
\providecommand \selectlanguage [0]{\@gobble}%
\providecommand \bibinfo  [0]{\@secondoftwo}%
\providecommand \bibfield  [0]{\@secondoftwo}%
\providecommand \translation [1]{[#1]}%
\providecommand \BibitemOpen [0]{}%
\providecommand \bibitemStop [0]{}%
\providecommand \bibitemNoStop [0]{.\EOS\space}%
\providecommand \EOS [0]{\spacefactor3000\relax}%
\providecommand \BibitemShut  [1]{\csname bibitem#1\endcsname}%
\let\auto@bib@innerbib\@empty
\bibitem [{\citenamefont {Abbott}\ \emph
  {et~al.}(2016{\natexlab{a}})\citenamefont {Abbott} \emph
  {et~al.}}]{GW150914-DETECTION}%
  \BibitemOpen
  \bibfield  {author} {\bibinfo {author} {\bibfnamefont {B.~P.}\ \bibnamefont
  {Abbott}} \emph {et~al.} (\bibinfo {collaboration} {LIGO Scientific
  Collaboration, Virgo Collaboration}),\ }\href@noop {} {\bibfield  {journal}
  {\bibinfo  {journal} {Phys.~Rev.~Lett.}\ }\textbf {\bibinfo {volume} {116}},\
  \bibinfo {pages} {061102} (\bibinfo {year} {2016}{\natexlab{a}})},\ \bibinfo
  {note} {\url{https://dcc.ligo.org/LIGO-P150914/public/main}},\ \Eprint
  {http://arxiv.org/abs/1602.03847} {arXiv:1602.03847 [gr-qc]} \BibitemShut
  {NoStop}%
\bibitem [{\citenamefont {Abbott}\ \emph
  {et~al.}(2016{\natexlab{b}})\citenamefont {Abbott} \emph
  {et~al.}}]{GW150914-PARAMESTIM}%
  \BibitemOpen
  \bibfield  {author} {\bibinfo {author} {\bibfnamefont {B.~P.}\ \bibnamefont
  {Abbott}} \emph {et~al.} (\bibinfo {collaboration} {LIGO Scientific
  Collaboration, Virgo Collaboration}),\ }\href@noop {} {\  (\bibinfo {year}
  {2016}{\natexlab{b}})},\ \bibinfo {note}
  {\url{https://dcc.ligo.org/LIGO-P1500218/public/main}},\ \Eprint
  {http://arxiv.org/abs/1602.03840} {arXiv:1602.03840 [gr-qc]} \BibitemShut
  {NoStop}%
\bibitem [{\citenamefont {Bruegmann}\ \emph {et~al.}(2008)\citenamefont
  {Bruegmann}, \citenamefont {Gonzalez}, \citenamefont {Hannam}, \citenamefont
  {Husa}, \citenamefont {Sperhake},\ and\ \citenamefont
  {Tichy}}]{Bruegmann:2006at}%
  \BibitemOpen
  \bibfield  {author} {\bibinfo {author} {\bibfnamefont {B.}~\bibnamefont
  {Bruegmann}}, \bibinfo {author} {\bibfnamefont {J.~A.}\ \bibnamefont
  {Gonzalez}}, \bibinfo {author} {\bibfnamefont {M.}~\bibnamefont {Hannam}},
  \bibinfo {author} {\bibfnamefont {S.}~\bibnamefont {Husa}}, \bibinfo {author}
  {\bibfnamefont {U.}~\bibnamefont {Sperhake}}, \ and\ \bibinfo {author}
  {\bibfnamefont {W.}~\bibnamefont {Tichy}},\ }\href {\doibase
  10.1103/PhysRevD.77.024027} {\bibfield  {journal} {\bibinfo  {journal} {Phys.
  Rev.}\ }\textbf {\bibinfo {volume} {D77}},\ \bibinfo {pages} {024027}
  (\bibinfo {year} {2008})},\ \Eprint {http://arxiv.org/abs/gr-qc/0610128}
  {arXiv:gr-qc/0610128 [gr-qc]} \BibitemShut {NoStop}%
\bibitem [{\citenamefont {O'Shaughnessy}\ \emph {et~al.}(2013)\citenamefont
  {O'Shaughnessy}, \citenamefont {London}, \citenamefont {Healy},\ and\
  \citenamefont {Shoemaker}}]{O'Shaughnessy:2012ay}%
  \BibitemOpen
  \bibfield  {author} {\bibinfo {author} {\bibfnamefont {R.}~\bibnamefont
  {O'Shaughnessy}}, \bibinfo {author} {\bibfnamefont {L.}~\bibnamefont
  {London}}, \bibinfo {author} {\bibfnamefont {J.}~\bibnamefont {Healy}}, \
  and\ \bibinfo {author} {\bibfnamefont {D.}~\bibnamefont {Shoemaker}},\ }\href
  {\doibase 10.1103/PhysRevD.87.044038} {\bibfield  {journal} {\bibinfo
  {journal} {Phys. Rev.}\ }\textbf {\bibinfo {volume} {D87}},\ \bibinfo {pages}
  {044038} (\bibinfo {year} {2013})},\ \Eprint {http://arxiv.org/abs/1209.3712}
  {arXiv:1209.3712 [gr-qc]} \BibitemShut {NoStop}%
\bibitem [{\citenamefont {Scheel}\ \emph {et~al.}(2015)\citenamefont {Scheel},
  \citenamefont {Giesler}, \citenamefont {Hemberger}, \citenamefont {Lovelace},
  \citenamefont {Kuper}, \citenamefont {Boyle}, \citenamefont {Szil{\'a}gyi},\
  and\ \citenamefont {Kidder}}]{Scheel:2014ina}%
  \BibitemOpen
  \bibfield  {author} {\bibinfo {author} {\bibfnamefont {M.~A.}\ \bibnamefont
  {Scheel}}, \bibinfo {author} {\bibfnamefont {M.}~\bibnamefont {Giesler}},
  \bibinfo {author} {\bibfnamefont {D.~A.}\ \bibnamefont {Hemberger}}, \bibinfo
  {author} {\bibfnamefont {G.}~\bibnamefont {Lovelace}}, \bibinfo {author}
  {\bibfnamefont {K.}~\bibnamefont {Kuper}}, \bibinfo {author} {\bibfnamefont
  {M.}~\bibnamefont {Boyle}}, \bibinfo {author} {\bibfnamefont
  {B.}~\bibnamefont {Szil{\'a}gyi}}, \ and\ \bibinfo {author} {\bibfnamefont
  {L.~E.}\ \bibnamefont {Kidder}},\ }\href {\doibase
  10.1088/0264-9381/32/10/105009} {\bibfield  {journal} {\bibinfo  {journal}
  {Class. Quant. Grav.}\ }\textbf {\bibinfo {volume} {32}},\ \bibinfo {pages}
  {105009} (\bibinfo {year} {2015})},\ \Eprint {http://arxiv.org/abs/1412.1803}
  {arXiv:1412.1803 [gr-qc]} \BibitemShut {NoStop}%
\bibitem [{\citenamefont {Chu}\ \emph {et~al.}(2015)\citenamefont {Chu},
  \citenamefont {Fong}, \citenamefont {Kumar}, \citenamefont {Pfeiffer},
  \citenamefont {Boyle}, \citenamefont {Hemberger}, \citenamefont {Kidder},
  \citenamefont {Scheel},\ and\ \citenamefont {Szil{\'a}gyi}}]{Chu:2015kft}%
  \BibitemOpen
  \bibfield  {author} {\bibinfo {author} {\bibfnamefont {T.}~\bibnamefont
  {Chu}}, \bibinfo {author} {\bibfnamefont {H.}~\bibnamefont {Fong}}, \bibinfo
  {author} {\bibfnamefont {P.}~\bibnamefont {Kumar}}, \bibinfo {author}
  {\bibfnamefont {H.~P.}\ \bibnamefont {Pfeiffer}}, \bibinfo {author}
  {\bibfnamefont {M.}~\bibnamefont {Boyle}}, \bibinfo {author} {\bibfnamefont
  {D.~A.}\ \bibnamefont {Hemberger}}, \bibinfo {author} {\bibfnamefont {L.~E.}\
  \bibnamefont {Kidder}}, \bibinfo {author} {\bibfnamefont {M.~A.}\
  \bibnamefont {Scheel}}, \ and\ \bibinfo {author} {\bibfnamefont
  {B.}~\bibnamefont {Szil{\'a}gyi}},\ }\href@noop {} {\enquote {\bibinfo
  {title} {{On the accuracy and precision of numerical waveforms: Effect of
  waveform extraction methodology}},}\ } (\bibinfo {year} {2015}),\ \bibinfo
  {note} {arXiv:1512.06800}\BibitemShut {NoStop}%
\bibitem [{\citenamefont {Lousto}\ \emph {et~al.}(2015)\citenamefont {Lousto},
  \citenamefont {Healy},\ and\ \citenamefont {Nakano}}]{Lousto:2015uwa}%
  \BibitemOpen
  \bibfield  {author} {\bibinfo {author} {\bibfnamefont {C.~O.}\ \bibnamefont
  {Lousto}}, \bibinfo {author} {\bibfnamefont {J.}~\bibnamefont {Healy}}, \
  and\ \bibinfo {author} {\bibfnamefont {H.}~\bibnamefont {Nakano}},\
  }\href@noop {} {\enquote {\bibinfo {title} {{Spin flips in generic black hole
  binaries}},}\ } (\bibinfo {year} {2015}),\ \bibinfo {note}
  {arXiv:1508.04768}\BibitemShut {NoStop}%
\bibitem [{\citenamefont {Szil{\'a}gyi}\ \emph {et~al.}(2015)\citenamefont
  {Szil{\'a}gyi}, \citenamefont {Blackman}, \citenamefont {Buonanno},
  \citenamefont {Taracchini}, \citenamefont {Pfeiffer}, \citenamefont {Scheel},
  \citenamefont {Chu}, \citenamefont {Kidder},\ and\ \citenamefont
  {Pan}}]{Szilagyi:2015rwa}%
  \BibitemOpen
  \bibfield  {author} {\bibinfo {author} {\bibfnamefont {B.}~\bibnamefont
  {Szil{\'a}gyi}}, \bibinfo {author} {\bibfnamefont {J.}~\bibnamefont
  {Blackman}}, \bibinfo {author} {\bibfnamefont {A.}~\bibnamefont {Buonanno}},
  \bibinfo {author} {\bibfnamefont {A.}~\bibnamefont {Taracchini}}, \bibinfo
  {author} {\bibfnamefont {H.~P.}\ \bibnamefont {Pfeiffer}}, \bibinfo {author}
  {\bibfnamefont {M.~A.}\ \bibnamefont {Scheel}}, \bibinfo {author}
  {\bibfnamefont {T.}~\bibnamefont {Chu}}, \bibinfo {author} {\bibfnamefont
  {L.~E.}\ \bibnamefont {Kidder}}, \ and\ \bibinfo {author} {\bibfnamefont
  {Y.}~\bibnamefont {Pan}},\ }\href {\doibase 10.1103/PhysRevLett.115.031102}
  {\bibfield  {journal} {\bibinfo  {journal} {Phys. Rev. Lett.}\ }\textbf
  {\bibinfo {volume} {115}},\ \bibinfo {pages} {031102} (\bibinfo {year}
  {2015})},\ \Eprint {http://arxiv.org/abs/1502.04953} {arXiv:1502.04953
  [gr-qc]} \BibitemShut {NoStop}%
\bibitem [{\citenamefont {Kumar}\ \emph {et~al.}(2015)\citenamefont {Kumar},
  \citenamefont {Barkett}, \citenamefont {Bhagwat}, \citenamefont {Afshari},
  \citenamefont {Brown}, \citenamefont {Lovelace}, \citenamefont {Scheel},\
  and\ \citenamefont {Szilágyi}}]{Kumar:2015tha}%
  \BibitemOpen
  \bibfield  {author} {\bibinfo {author} {\bibfnamefont {P.}~\bibnamefont
  {Kumar}}, \bibinfo {author} {\bibfnamefont {K.}~\bibnamefont {Barkett}},
  \bibinfo {author} {\bibfnamefont {S.}~\bibnamefont {Bhagwat}}, \bibinfo
  {author} {\bibfnamefont {N.}~\bibnamefont {Afshari}}, \bibinfo {author}
  {\bibfnamefont {D.~A.}\ \bibnamefont {Brown}}, \bibinfo {author}
  {\bibfnamefont {G.}~\bibnamefont {Lovelace}}, \bibinfo {author}
  {\bibfnamefont {M.~A.}\ \bibnamefont {Scheel}}, \ and\ \bibinfo {author}
  {\bibfnamefont {B.}~\bibnamefont {Szilágyi}},\ }\href {\doibase
  10.1103/PhysRevD.92.102001} {\bibfield  {journal} {\bibinfo  {journal} {Phys.
  Rev.}\ }\textbf {\bibinfo {volume} {D92}},\ \bibinfo {pages} {102001}
  (\bibinfo {year} {2015})},\ \Eprint {http://arxiv.org/abs/1507.00103}
  {arXiv:1507.00103 [gr-qc]} \BibitemShut {NoStop}%
\bibitem [{\citenamefont {Lousto}\ and\ \citenamefont
  {Healy}(2015)}]{Lousto:2014ida}%
  \BibitemOpen
  \bibfield  {author} {\bibinfo {author} {\bibfnamefont {C.~O.}\ \bibnamefont
  {Lousto}}\ and\ \bibinfo {author} {\bibfnamefont {J.}~\bibnamefont {Healy}},\
  }\href {\doibase 10.1103/PhysRevLett.114.141101} {\bibfield  {journal}
  {\bibinfo  {journal} {Phys. Rev. Lett.}\ }\textbf {\bibinfo {volume} {114}},\
  \bibinfo {pages} {141101} (\bibinfo {year} {2015})},\ \Eprint
  {http://arxiv.org/abs/1410.3830} {arXiv:1410.3830 [gr-qc]} \BibitemShut
  {NoStop}%
\bibitem [{\citenamefont {Buonanno}\ and\ \citenamefont
  {Damour}(1999)}]{Buonanno:1998gg}%
  \BibitemOpen
  \bibfield  {author} {\bibinfo {author} {\bibfnamefont {A.}~\bibnamefont
  {Buonanno}}\ and\ \bibinfo {author} {\bibfnamefont {T.}~\bibnamefont
  {Damour}},\ }\href {\doibase 10.1103/PhysRevD.59.084006} {\bibfield
  {journal} {\bibinfo  {journal} {Phys. Rev.}\ }\textbf {\bibinfo {volume}
  {D59}},\ \bibinfo {pages} {084006} (\bibinfo {year} {1999})},\ \Eprint
  {http://arxiv.org/abs/gr-qc/9811091} {arXiv:gr-qc/9811091 [gr-qc]}
  \BibitemShut {NoStop}%
\bibitem [{\citenamefont {Buonanno}\ and\ \citenamefont
  {Damour}(2000)}]{Buonanno:2000ef}%
  \BibitemOpen
  \bibfield  {author} {\bibinfo {author} {\bibfnamefont {A.}~\bibnamefont
  {Buonanno}}\ and\ \bibinfo {author} {\bibfnamefont {T.}~\bibnamefont
  {Damour}},\ }\href {\doibase 10.1103/PhysRevD.62.064015} {\bibfield
  {journal} {\bibinfo  {journal} {Phys. Rev.}\ }\textbf {\bibinfo {volume}
  {D62}},\ \bibinfo {pages} {064015} (\bibinfo {year} {2000})},\ \Eprint
  {http://arxiv.org/abs/gr-qc/0001013} {arXiv:gr-qc/0001013 [gr-qc]}
  \BibitemShut {NoStop}%
\bibitem [{\citenamefont {Taracchini}\ \emph
  {et~al.}(2014{\natexlab{a}})\citenamefont {Taracchini} \emph
  {et~al.}}]{Taracchini:2013rva}%
  \BibitemOpen
  \bibfield  {author} {\bibinfo {author} {\bibfnamefont {A.}~\bibnamefont
  {Taracchini}} \emph {et~al.},\ }\href {\doibase 10.1103/PhysRevD.89.061502}
  {\bibfield  {journal} {\bibinfo  {journal} {Phys. Rev.}\ }\textbf {\bibinfo
  {volume} {D89}},\ \bibinfo {pages} {061502} (\bibinfo {year}
  {2014}{\natexlab{a}})},\ \Eprint {http://arxiv.org/abs/1311.2544}
  {arXiv:1311.2544 [gr-qc]} \BibitemShut {NoStop}%
\bibitem [{\citenamefont {Hannam}(2014)}]{Hannam:2013pra}%
  \BibitemOpen
  \bibfield  {author} {\bibinfo {author} {\bibfnamefont {M.}~\bibnamefont
  {Hannam}},\ }\href {\doibase 10.1007/s10714-014-1767-2} {\bibfield  {journal}
  {\bibinfo  {journal} {Gen. Rel. Grav.}\ }\textbf {\bibinfo {volume} {46}},\
  \bibinfo {pages} {1767} (\bibinfo {year} {2014})},\ \Eprint
  {http://arxiv.org/abs/1312.3641} {arXiv:1312.3641 [gr-qc]} \BibitemShut
  {NoStop}%
\bibitem [{\citenamefont {Pan}\ \emph {et~al.}(2014)\citenamefont {Pan},
  \citenamefont {Buonanno}, \citenamefont {Taracchini}, \citenamefont {Kidder},
  \citenamefont {Mrou{\'e}}, \citenamefont {Pfeiffer}, \citenamefont {Scheel},\
  and\ \citenamefont {Szilágyi}}]{Pan:2013rra}%
  \BibitemOpen
  \bibfield  {author} {\bibinfo {author} {\bibfnamefont {Y.}~\bibnamefont
  {Pan}}, \bibinfo {author} {\bibfnamefont {A.}~\bibnamefont {Buonanno}},
  \bibinfo {author} {\bibfnamefont {A.}~\bibnamefont {Taracchini}}, \bibinfo
  {author} {\bibfnamefont {L.~E.}\ \bibnamefont {Kidder}}, \bibinfo {author}
  {\bibfnamefont {A.~H.}\ \bibnamefont {Mrou{\'e}}}, \bibinfo {author}
  {\bibfnamefont {H.~P.}\ \bibnamefont {Pfeiffer}}, \bibinfo {author}
  {\bibfnamefont {M.~A.}\ \bibnamefont {Scheel}}, \ and\ \bibinfo {author}
  {\bibfnamefont {B.}~\bibnamefont {Szilágyi}},\ }\href {\doibase
  10.1103/PhysRevD.89.084006} {\bibfield  {journal} {\bibinfo  {journal} {Phys.
  Rev.}\ }\textbf {\bibinfo {volume} {D89}},\ \bibinfo {pages} {084006}
  (\bibinfo {year} {2014})},\ \Eprint {http://arxiv.org/abs/1307.6232}
  {arXiv:1307.6232 [gr-qc]} \BibitemShut {NoStop}%
\bibitem [{\citenamefont {Babak}\ \emph {et~al.}(2016)\citenamefont {Babak},
  \citenamefont {Taracchini},\ and\ \citenamefont
  {Buonanno}}]{BabaketalInPrep}%
  \BibitemOpen
  \bibfield  {author} {\bibinfo {author} {\bibfnamefont {S.}~\bibnamefont
  {Babak}}, \bibinfo {author} {\bibfnamefont {A.}~\bibnamefont {Taracchini}}, \
  and\ \bibinfo {author} {\bibfnamefont {A.}~\bibnamefont {Buonanno}},\
  }\href@noop {} {\bibfield  {journal} {\bibinfo  {journal} {In preparation}\ }
  (\bibinfo {year} {2016})}\BibitemShut {NoStop}%
\bibitem [{\citenamefont {Aasi}\ \emph {et~al.}(2013)\citenamefont {Aasi} \emph
  {et~al.}}]{S6pepaper}%
  \BibitemOpen
  \bibfield  {author} {\bibinfo {author} {\bibfnamefont {J.}~\bibnamefont
  {Aasi}} \emph {et~al.} (\bibinfo {collaboration} {LIGO Collaboration, Virgo
  Collaboration}),\ }\href {\doibase 10.1103/PhysRevD.88.062001} {\bibfield
  {journal} {\bibinfo  {journal} {Phys.Rev.}\ }\textbf {\bibinfo {volume}
  {D88}},\ \bibinfo {pages} {062001} (\bibinfo {year} {2013})},\ \Eprint
  {http://arxiv.org/abs/1304.1775} {arXiv:1304.1775 [gr-qc]} \BibitemShut
  {NoStop}%
\bibitem [{\citenamefont {Abbott}\ \emph
  {et~al.}(2016{\natexlab{c}})\citenamefont {Abbott} \emph
  {et~al.}}]{GW150914-ASTRO}%
  \BibitemOpen
  \bibfield  {author} {\bibinfo {author} {\bibfnamefont {B.~P.}\ \bibnamefont
  {Abbott}} \emph {et~al.} (\bibinfo {collaboration} {LIGO Scientific
  Collaboration, Virgo Collaboration}),\ }\href {\doibase
  10.3847/2041-8205/818/2/L22} {\bibfield  {journal} {\bibinfo  {journal}
  {Astrophys.~J.}\ }\textbf {\bibinfo {volume} {818}},\ \bibinfo {pages} {L22}
  (\bibinfo {year} {2016}{\natexlab{c}})},\ \bibinfo {note}
  {\url{https://dcc.ligo.org/LIGO-P1500262/public/main}},\ \Eprint
  {http://arxiv.org/abs/1602.03846} {arXiv:1602.03846 [astro-ph.HE]}
  \BibitemShut {NoStop}%
\bibitem [{\citenamefont {Damour}(2001)}]{Damour:2001tu}%
  \BibitemOpen
  \bibfield  {author} {\bibinfo {author} {\bibfnamefont {T.}~\bibnamefont
  {Damour}},\ }\href {\doibase 10.1103/PhysRevD.64.124013} {\bibfield
  {journal} {\bibinfo  {journal} {Phys. Rev.}\ }\textbf {\bibinfo {volume}
  {D64}},\ \bibinfo {pages} {124013} (\bibinfo {year} {2001})},\ \Eprint
  {http://arxiv.org/abs/gr-qc/0103018} {arXiv:gr-qc/0103018 [gr-qc]}
  \BibitemShut {NoStop}%
\bibitem [{\citenamefont {Buonanno}\ \emph {et~al.}(2006)\citenamefont
  {Buonanno}, \citenamefont {Chen},\ and\ \citenamefont
  {Damour}}]{Buonanno:2005xu}%
  \BibitemOpen
  \bibfield  {author} {\bibinfo {author} {\bibfnamefont {A.}~\bibnamefont
  {Buonanno}}, \bibinfo {author} {\bibfnamefont {Y.}~\bibnamefont {Chen}}, \
  and\ \bibinfo {author} {\bibfnamefont {T.}~\bibnamefont {Damour}},\ }\href
  {\doibase 10.1103/PhysRevD.74.104005} {\bibfield  {journal} {\bibinfo
  {journal} {Phys. Rev.}\ }\textbf {\bibinfo {volume} {D74}},\ \bibinfo {pages}
  {104005} (\bibinfo {year} {2006})},\ \Eprint
  {http://arxiv.org/abs/gr-qc/0508067} {arXiv:gr-qc/0508067 [gr-qc]}
  \BibitemShut {NoStop}%
\bibitem [{\citenamefont {Damour}\ \emph {et~al.}(2008)\citenamefont {Damour},
  \citenamefont {Jaranowski},\ and\ \citenamefont {Schaefer}}]{Damour:2008qf}%
  \BibitemOpen
  \bibfield  {author} {\bibinfo {author} {\bibfnamefont {T.}~\bibnamefont
  {Damour}}, \bibinfo {author} {\bibfnamefont {P.}~\bibnamefont {Jaranowski}},
  \ and\ \bibinfo {author} {\bibfnamefont {G.}~\bibnamefont {Schaefer}},\
  }\href {\doibase 10.1103/PhysRevD.78.024009} {\bibfield  {journal} {\bibinfo
  {journal} {Phys. Rev.}\ }\textbf {\bibinfo {volume} {D78}},\ \bibinfo {pages}
  {024009} (\bibinfo {year} {2008})},\ \Eprint {http://arxiv.org/abs/0803.0915}
  {arXiv:0803.0915 [gr-qc]} \BibitemShut {NoStop}%
\bibitem [{\citenamefont {Barausse}\ and\ \citenamefont
  {Buonanno}(2010)}]{Barausse:2009xi}%
  \BibitemOpen
  \bibfield  {author} {\bibinfo {author} {\bibfnamefont {E.}~\bibnamefont
  {Barausse}}\ and\ \bibinfo {author} {\bibfnamefont {A.}~\bibnamefont
  {Buonanno}},\ }\href {\doibase 10.1103/PhysRevD.81.084024} {\bibfield
  {journal} {\bibinfo  {journal} {Phys. Rev.}\ }\textbf {\bibinfo {volume}
  {D81}},\ \bibinfo {pages} {084024} (\bibinfo {year} {2010})},\ \Eprint
  {http://arxiv.org/abs/0912.3517} {arXiv:0912.3517 [gr-qc]} \BibitemShut
  {NoStop}%
\bibitem [{\citenamefont {Barausse}\ and\ \citenamefont
  {Buonanno}(2011)}]{Barausse:2011ys}%
  \BibitemOpen
  \bibfield  {author} {\bibinfo {author} {\bibfnamefont {E.}~\bibnamefont
  {Barausse}}\ and\ \bibinfo {author} {\bibfnamefont {A.}~\bibnamefont
  {Buonanno}},\ }\href {\doibase 10.1103/PhysRevD.84.104027} {\bibfield
  {journal} {\bibinfo  {journal} {Phys. Rev.}\ }\textbf {\bibinfo {volume}
  {D84}},\ \bibinfo {pages} {104027} (\bibinfo {year} {2011})},\ \Eprint
  {http://arxiv.org/abs/1107.2904} {arXiv:1107.2904 [gr-qc]} \BibitemShut
  {NoStop}%
\bibitem [{\citenamefont {Damour}\ and\ \citenamefont
  {Nagar}(2014)}]{Damour:2014sva}%
  \BibitemOpen
  \bibfield  {author} {\bibinfo {author} {\bibfnamefont {T.}~\bibnamefont
  {Damour}}\ and\ \bibinfo {author} {\bibfnamefont {A.}~\bibnamefont {Nagar}},\
  }\href {\doibase 10.1103/PhysRevD.90.044018} {\bibfield  {journal} {\bibinfo
  {journal} {Phys. Rev.}\ }\textbf {\bibinfo {volume} {D90}},\ \bibinfo {pages}
  {044018} (\bibinfo {year} {2014})},\ \Eprint {http://arxiv.org/abs/1406.6913}
  {arXiv:1406.6913 [gr-qc]} \BibitemShut {NoStop}%
\bibitem [{\citenamefont {Blanchet}(2014)}]{blanchet:2014}%
  \BibitemOpen
  \bibfield  {author} {\bibinfo {author} {\bibfnamefont {L.}~\bibnamefont
  {Blanchet}},\ }\href@noop {} {\bibfield  {journal} {\bibinfo  {journal}
  {Living Rev. Rel.}\ }\textbf {\bibinfo {volume} {17}},\ \bibinfo {pages} {2}
  (\bibinfo {year} {2014})}\BibitemShut {NoStop}%
\bibitem [{\citenamefont {Damour}\ and\ \citenamefont
  {Nagar}(2007)}]{Damour:2007xr}%
  \BibitemOpen
  \bibfield  {author} {\bibinfo {author} {\bibfnamefont {T.}~\bibnamefont
  {Damour}}\ and\ \bibinfo {author} {\bibfnamefont {A.}~\bibnamefont {Nagar}},\
  }\href {\doibase 10.1103/PhysRevD.76.064028} {\bibfield  {journal} {\bibinfo
  {journal} {Phys. Rev.}\ }\textbf {\bibinfo {volume} {D76}},\ \bibinfo {pages}
  {064028} (\bibinfo {year} {2007})},\ \Eprint {http://arxiv.org/abs/0705.2519}
  {arXiv:0705.2519 [gr-qc]} \BibitemShut {NoStop}%
\bibitem [{\citenamefont {Damour}\ \emph {et~al.}(2009)\citenamefont {Damour},
  \citenamefont {Iyer},\ and\ \citenamefont {Nagar}}]{Damour:2008gu}%
  \BibitemOpen
  \bibfield  {author} {\bibinfo {author} {\bibfnamefont {T.}~\bibnamefont
  {Damour}}, \bibinfo {author} {\bibfnamefont {B.~R.}\ \bibnamefont {Iyer}}, \
  and\ \bibinfo {author} {\bibfnamefont {A.}~\bibnamefont {Nagar}},\ }\href
  {\doibase 10.1103/PhysRevD.79.064004} {\bibfield  {journal} {\bibinfo
  {journal} {Phys. Rev.}\ }\textbf {\bibinfo {volume} {D79}},\ \bibinfo {pages}
  {064004} (\bibinfo {year} {2009})},\ \Eprint {http://arxiv.org/abs/0811.2069}
  {arXiv:0811.2069 [gr-qc]} \BibitemShut {NoStop}%
\bibitem [{\citenamefont {Pan}\ \emph {et~al.}(2011)\citenamefont {Pan},
  \citenamefont {Buonanno}, \citenamefont {Fujita}, \citenamefont {Racine},\
  and\ \citenamefont {Tagoshi}}]{Pan:2010hz}%
  \BibitemOpen
  \bibfield  {author} {\bibinfo {author} {\bibfnamefont {Y.}~\bibnamefont
  {Pan}}, \bibinfo {author} {\bibfnamefont {A.}~\bibnamefont {Buonanno}},
  \bibinfo {author} {\bibfnamefont {R.}~\bibnamefont {Fujita}}, \bibinfo
  {author} {\bibfnamefont {E.}~\bibnamefont {Racine}}, \ and\ \bibinfo {author}
  {\bibfnamefont {H.}~\bibnamefont {Tagoshi}},\ }\href {\doibase
  10.1103/PhysRevD.83.064003, 10.1103/PhysRevD.87.109901} {\bibfield  {journal}
  {\bibinfo  {journal} {Phys. Rev.}\ }\textbf {\bibinfo {volume} {D83}},\
  \bibinfo {pages} {064003} (\bibinfo {year} {2011})},\ \bibinfo {note}
  {[Erratum: Phys. Rev.D87,no.10,109901(2013)]},\ \Eprint
  {http://arxiv.org/abs/1006.0431} {arXiv:1006.0431 [gr-qc]} \BibitemShut
  {NoStop}%
\bibitem [{\citenamefont {Vishveshwara}(1970)}]{Vishveshwara:1970zz}%
  \BibitemOpen
  \bibfield  {author} {\bibinfo {author} {\bibfnamefont {C.~V.}\ \bibnamefont
  {Vishveshwara}},\ }\href@noop {} {\bibfield  {journal} {\bibinfo  {journal}
  {Nature}\ }\textbf {\bibinfo {volume} {227}},\ \bibinfo {pages} {936}
  (\bibinfo {year} {1970})}\BibitemShut {NoStop}%
\bibitem [{\citenamefont {Press}(1971)}]{Press:1971wr}%
  \BibitemOpen
  \bibfield  {author} {\bibinfo {author} {\bibfnamefont {W.~H.}\ \bibnamefont
  {Press}},\ }\href@noop {} {\bibfield  {journal} {\bibinfo  {journal}
  {Astrophys. J.}\ }\textbf {\bibinfo {volume} {170}},\ \bibinfo {pages} {L105}
  (\bibinfo {year} {1971})}\BibitemShut {NoStop}%
\bibitem [{\citenamefont {Chandrasekhar}\ and\ \citenamefont
  {Detweiler}(1975)}]{Chandrasekhar:1975zza}%
  \BibitemOpen
  \bibfield  {author} {\bibinfo {author} {\bibfnamefont {S.}~\bibnamefont
  {Chandrasekhar}}\ and\ \bibinfo {author} {\bibfnamefont {S.~L.}\ \bibnamefont
  {Detweiler}},\ }\href@noop {} {\bibfield  {journal} {\bibinfo  {journal}
  {Proc. Roy. Soc. Lond.}\ }\textbf {\bibinfo {volume} {A344}},\ \bibinfo
  {pages} {441} (\bibinfo {year} {1975})}\BibitemShut {NoStop}%
\bibitem [{\citenamefont {Mrou{\'e}}\ \emph {et~al.}(2013)\citenamefont
  {Mrou{\'e}} \emph {et~al.}}]{Mroue:2013xna}%
  \BibitemOpen
  \bibfield  {author} {\bibinfo {author} {\bibfnamefont {A.~H.}\ \bibnamefont
  {Mrou{\'e}}} \emph {et~al.},\ }\href {\doibase
  10.1103/PhysRevLett.111.241104} {\bibfield  {journal} {\bibinfo  {journal}
  {Phys. Rev. Lett.}\ }\textbf {\bibinfo {volume} {111}},\ \bibinfo {pages}
  {241104} (\bibinfo {year} {2013})},\ \Eprint {http://arxiv.org/abs/1304.6077}
  {arXiv:1304.6077 [gr-qc]} \BibitemShut {NoStop}%
\bibitem [{\citenamefont {Barausse}\ \emph {et~al.}(2012)\citenamefont
  {Barausse}, \citenamefont {Buonanno}, \citenamefont {Hughes}, \citenamefont
  {Khanna}, \citenamefont {O'Sullivan},\ and\ \citenamefont
  {Pan}}]{Barausse:2011kb}%
  \BibitemOpen
  \bibfield  {author} {\bibinfo {author} {\bibfnamefont {E.}~\bibnamefont
  {Barausse}}, \bibinfo {author} {\bibfnamefont {A.}~\bibnamefont {Buonanno}},
  \bibinfo {author} {\bibfnamefont {S.~A.}\ \bibnamefont {Hughes}}, \bibinfo
  {author} {\bibfnamefont {G.}~\bibnamefont {Khanna}}, \bibinfo {author}
  {\bibfnamefont {S.}~\bibnamefont {O'Sullivan}}, \ and\ \bibinfo {author}
  {\bibfnamefont {Y.}~\bibnamefont {Pan}},\ }\href {\doibase
  10.1103/PhysRevD.85.024046} {\bibfield  {journal} {\bibinfo  {journal} {Phys.
  Rev.}\ }\textbf {\bibinfo {volume} {D85}},\ \bibinfo {pages} {024046}
  (\bibinfo {year} {2012})},\ \Eprint {http://arxiv.org/abs/1110.3081}
  {arXiv:1110.3081 [gr-qc]} \BibitemShut {NoStop}%
\bibitem [{\citenamefont {Taracchini}\ \emph
  {et~al.}(2014{\natexlab{b}})\citenamefont {Taracchini}, \citenamefont
  {Buonanno}, \citenamefont {Khanna},\ and\ \citenamefont
  {Hughes}}]{Taracchini:2014zpa}%
  \BibitemOpen
  \bibfield  {author} {\bibinfo {author} {\bibfnamefont {A.}~\bibnamefont
  {Taracchini}}, \bibinfo {author} {\bibfnamefont {A.}~\bibnamefont
  {Buonanno}}, \bibinfo {author} {\bibfnamefont {G.}~\bibnamefont {Khanna}}, \
  and\ \bibinfo {author} {\bibfnamefont {S.~A.}\ \bibnamefont {Hughes}},\
  }\href {\doibase 10.1103/PhysRevD.90.084025} {\bibfield  {journal} {\bibinfo
  {journal} {Phys. Rev.}\ }\textbf {\bibinfo {volume} {D90}},\ \bibinfo {pages}
  {084025} (\bibinfo {year} {2014}{\natexlab{b}})},\ \Eprint
  {http://arxiv.org/abs/1404.1819} {arXiv:1404.1819 [gr-qc]} \BibitemShut
  {NoStop}%
\bibitem [{\citenamefont {Buonanno}\ \emph {et~al.}(2004)\citenamefont
  {Buonanno}, \citenamefont {Chen}, \citenamefont {Pan},\ and\ \citenamefont
  {Vallisneri}}]{buonanno:2004yd}%
  \BibitemOpen
  \bibfield  {author} {\bibinfo {author} {\bibfnamefont {A.}~\bibnamefont
  {Buonanno}}, \bibinfo {author} {\bibfnamefont {Y.-b.}\ \bibnamefont {Chen}},
  \bibinfo {author} {\bibfnamefont {Y.}~\bibnamefont {Pan}}, \ and\ \bibinfo
  {author} {\bibfnamefont {M.}~\bibnamefont {Vallisneri}},\ }\href@noop {}
  {\bibfield  {journal} {\bibinfo  {journal} {\prd}\ }\textbf {\bibinfo
  {volume} {70}},\ \bibinfo {pages} {104003} (\bibinfo {year} {2004})},\
  \bibinfo {note} {erratum-ibid. 74 (2006) 029902(E)},\ \Eprint
  {http://arxiv.org/abs/gr-qc/0405090} {gr-qc/0405090} \BibitemShut {NoStop}%
\bibitem [{\citenamefont {Boyle}\ \emph {et~al.}(2011)\citenamefont {Boyle},
  \citenamefont {Owen},\ and\ \citenamefont {Pfeiffer}}]{Boyle:2011gg}%
  \BibitemOpen
  \bibfield  {author} {\bibinfo {author} {\bibfnamefont {M.}~\bibnamefont
  {Boyle}}, \bibinfo {author} {\bibfnamefont {R.}~\bibnamefont {Owen}}, \ and\
  \bibinfo {author} {\bibfnamefont {H.~P.}\ \bibnamefont {Pfeiffer}},\ }\href
  {\doibase 10.1103/PhysRevD.84.124011} {\bibfield  {journal} {\bibinfo
  {journal} {Phys. Rev.}\ }\textbf {\bibinfo {volume} {D84}},\ \bibinfo {pages}
  {124011} (\bibinfo {year} {2011})},\ \Eprint {http://arxiv.org/abs/1110.2965}
  {arXiv:1110.2965 [gr-qc]} \BibitemShut {NoStop}%
\bibitem [{\citenamefont {Schmidt}\ \emph {et~al.}(2011)\citenamefont
  {Schmidt}, \citenamefont {Hannam}, \citenamefont {Husa},\ and\ \citenamefont
  {Ajith}}]{Schmidt:2010it}%
  \BibitemOpen
  \bibfield  {author} {\bibinfo {author} {\bibfnamefont {P.}~\bibnamefont
  {Schmidt}}, \bibinfo {author} {\bibfnamefont {M.}~\bibnamefont {Hannam}},
  \bibinfo {author} {\bibfnamefont {S.}~\bibnamefont {Husa}}, \ and\ \bibinfo
  {author} {\bibfnamefont {P.}~\bibnamefont {Ajith}},\ }\href {\doibase
  10.1103/PhysRevD.84.024046} {\bibfield  {journal} {\bibinfo  {journal} {Phys.
  Rev.}\ }\textbf {\bibinfo {volume} {D84}},\ \bibinfo {pages} {024046}
  (\bibinfo {year} {2011})},\ \Eprint {http://arxiv.org/abs/1012.2879}
  {arXiv:1012.2879 [gr-qc]} \BibitemShut {NoStop}%
\bibitem [{\citenamefont {O'Shaughnessy}\ \emph {et~al.}(2011)\citenamefont
  {O'Shaughnessy}, \citenamefont {Vaishnav}, \citenamefont {Healy},
  \citenamefont {Meeks},\ and\ \citenamefont
  {Shoemaker}}]{O'Shaughnessy:2011fx}%
  \BibitemOpen
  \bibfield  {author} {\bibinfo {author} {\bibfnamefont {R.}~\bibnamefont
  {O'Shaughnessy}}, \bibinfo {author} {\bibfnamefont {B.}~\bibnamefont
  {Vaishnav}}, \bibinfo {author} {\bibfnamefont {J.}~\bibnamefont {Healy}},
  \bibinfo {author} {\bibfnamefont {Z.}~\bibnamefont {Meeks}}, \ and\ \bibinfo
  {author} {\bibfnamefont {D.}~\bibnamefont {Shoemaker}},\ }\href {\doibase
  10.1103/PhysRevD.84.124002} {\bibfield  {journal} {\bibinfo  {journal} {Phys.
  Rev.}\ }\textbf {\bibinfo {volume} {D84}},\ \bibinfo {pages} {124002}
  (\bibinfo {year} {2011})},\ \Eprint {http://arxiv.org/abs/1109.5224}
  {arXiv:1109.5224 [gr-qc]} \BibitemShut {NoStop}%
\bibitem [{\citenamefont {Schmidt}\ \emph {et~al.}(2012)\citenamefont
  {Schmidt}, \citenamefont {Hannam},\ and\ \citenamefont
  {Husa}}]{Schmidt:2012rh}%
  \BibitemOpen
  \bibfield  {author} {\bibinfo {author} {\bibfnamefont {P.}~\bibnamefont
  {Schmidt}}, \bibinfo {author} {\bibfnamefont {M.}~\bibnamefont {Hannam}}, \
  and\ \bibinfo {author} {\bibfnamefont {S.}~\bibnamefont {Husa}},\ }\href
  {\doibase 10.1103/PhysRevD.86.104063} {\bibfield  {journal} {\bibinfo
  {journal} {Phys. Rev.}\ }\textbf {\bibinfo {volume} {D86}},\ \bibinfo {pages}
  {104063} (\bibinfo {year} {2012})},\ \Eprint {http://arxiv.org/abs/1207.3088}
  {arXiv:1207.3088 [gr-qc]} \BibitemShut {NoStop}%
\bibitem [{\citenamefont {Schmidt}\ \emph {et~al.}(2015)\citenamefont
  {Schmidt}, \citenamefont {Ohme},\ and\ \citenamefont
  {Hannam}}]{Schmidt:2014iyl}%
  \BibitemOpen
  \bibfield  {author} {\bibinfo {author} {\bibfnamefont {P.}~\bibnamefont
  {Schmidt}}, \bibinfo {author} {\bibfnamefont {F.}~\bibnamefont {Ohme}}, \
  and\ \bibinfo {author} {\bibfnamefont {M.}~\bibnamefont {Hannam}},\ }\href
  {\doibase 10.1103/PhysRevD.91.024043} {\bibfield  {journal} {\bibinfo
  {journal} {Phys. Rev.}\ }\textbf {\bibinfo {volume} {D91}},\ \bibinfo {pages}
  {024043} (\bibinfo {year} {2015})},\ \Eprint {http://arxiv.org/abs/1408.1810}
  {arXiv:1408.1810 [gr-qc]} \BibitemShut {NoStop}%
\bibitem [{\citenamefont {Hannam}\ \emph {et~al.}(2014)\citenamefont {Hannam},
  \citenamefont {Schmidt}, \citenamefont {Boh{\'e}}, \citenamefont {Haegel},
  \citenamefont {Husa}, \citenamefont {Ohme}, \citenamefont {Pratten},\ and\
  \citenamefont {P{\"u}rrer}}]{Hannam:2013oca}%
  \BibitemOpen
  \bibfield  {author} {\bibinfo {author} {\bibfnamefont {M.}~\bibnamefont
  {Hannam}}, \bibinfo {author} {\bibfnamefont {P.}~\bibnamefont {Schmidt}},
  \bibinfo {author} {\bibfnamefont {A.}~\bibnamefont {Boh{\'e}}}, \bibinfo
  {author} {\bibfnamefont {L.}~\bibnamefont {Haegel}}, \bibinfo {author}
  {\bibfnamefont {S.}~\bibnamefont {Husa}}, \bibinfo {author} {\bibfnamefont
  {F.}~\bibnamefont {Ohme}}, \bibinfo {author} {\bibfnamefont {G.}~\bibnamefont
  {Pratten}}, \ and\ \bibinfo {author} {\bibfnamefont {M.}~\bibnamefont
  {P{\"u}rrer}},\ }\href {\doibase 10.1103/PhysRevLett.113.151101} {\bibfield
  {journal} {\bibinfo  {journal} {Phys. Rev. Lett.}\ }\textbf {\bibinfo
  {volume} {113}},\ \bibinfo {pages} {151101} (\bibinfo {year} {2014})},\
  \Eprint {http://arxiv.org/abs/1308.3271} {arXiv:1308.3271 [gr-qc]}
  \BibitemShut {NoStop}%
\bibitem [{\citenamefont {Khan}\ \emph {et~al.}(2015)\citenamefont {Khan},
  \citenamefont {Husa}, \citenamefont {Hannam}, \citenamefont {Ohme},
  \citenamefont {P{\"u}rrer}, \citenamefont {Forteza},\ and\ \citenamefont
  {Boh{\'e}}}]{khan:2015jqa}%
  \BibitemOpen
  \bibfield  {author} {\bibinfo {author} {\bibfnamefont {S.}~\bibnamefont
  {Khan}}, \bibinfo {author} {\bibfnamefont {S.}~\bibnamefont {Husa}}, \bibinfo
  {author} {\bibfnamefont {M.}~\bibnamefont {Hannam}}, \bibinfo {author}
  {\bibfnamefont {F.}~\bibnamefont {Ohme}}, \bibinfo {author} {\bibfnamefont
  {M.}~\bibnamefont {P{\"u}rrer}}, \bibinfo {author} {\bibfnamefont {X.~J.}\
  \bibnamefont {Forteza}}, \ and\ \bibinfo {author} {\bibfnamefont
  {A.}~\bibnamefont {Boh{\'e}}},\ }\href@noop {} {\enquote {\bibinfo {title}
  {{Frequency-domain gravitational waves from non-precessing black-hole
  binaries. II. A phenomenological model for the advanced detector era}},}\ }
  (\bibinfo {year} {2015}),\ \bibinfo {note} {arXiv:1508.07253}\BibitemShut
  {NoStop}%
\bibitem [{\citenamefont {Kumar}\ \emph {et~al.}(2016)\citenamefont {Kumar},
  \citenamefont {Chu}, \citenamefont {Fong}, \citenamefont {Pfeiffer},
  \citenamefont {Boyle}, \citenamefont {Hemberger}, \citenamefont {Kidder},
  \citenamefont {Scheel},\ and\ \citenamefont {Szil{\'a}gyi}}]{Kumar:2016dhh}%
  \BibitemOpen
  \bibfield  {author} {\bibinfo {author} {\bibfnamefont {P.}~\bibnamefont
  {Kumar}}, \bibinfo {author} {\bibfnamefont {T.}~\bibnamefont {Chu}}, \bibinfo
  {author} {\bibfnamefont {H.}~\bibnamefont {Fong}}, \bibinfo {author}
  {\bibfnamefont {H.~P.}\ \bibnamefont {Pfeiffer}}, \bibinfo {author}
  {\bibfnamefont {M.}~\bibnamefont {Boyle}}, \bibinfo {author} {\bibfnamefont
  {D.~A.}\ \bibnamefont {Hemberger}}, \bibinfo {author} {\bibfnamefont {L.~E.}\
  \bibnamefont {Kidder}}, \bibinfo {author} {\bibfnamefont {M.~A.}\
  \bibnamefont {Scheel}}, \ and\ \bibinfo {author} {\bibfnamefont
  {B.}~\bibnamefont {Szil{\'a}gyi}},\ }\href@noop {} {\enquote {\bibinfo
  {title} {{Accuracy of binary black hole waveform models for aligned-spin
  binaries}},}\ } (\bibinfo {year} {2016}),\ \bibinfo {note}
  {arXiv:1601.05396}\BibitemShut {NoStop}%
\bibitem [{\citenamefont {Bayes}\ and\ \citenamefont
  {Price}(1763)}]{Bayes:1793}%
  \BibitemOpen
  \bibfield  {author} {\bibinfo {author} {\bibfnamefont {T.}~\bibnamefont
  {Bayes}}\ and\ \bibinfo {author} {\bibfnamefont {R.}~\bibnamefont {Price}},\
  }\href {\doibase 10.1098/rstl.1763.0053} {\bibfield  {journal} {\bibinfo
  {journal} {Phil. Trans. Roy. Soc. Lond.}\ }\textbf {\bibinfo {volume} {53}},\
  \bibinfo {pages} {370} (\bibinfo {year} {1763})}\BibitemShut {NoStop}%
\bibitem [{\citenamefont {Jaynes}(2003)}]{Jaynes:2003}%
  \BibitemOpen
  \bibfield  {author} {\bibinfo {author} {\bibfnamefont {E.~T.}\ \bibnamefont
  {Jaynes}},\ }\href@noop {} {\emph {\bibinfo {title} {{Probability Theory: The
  Logic of Science}}}},\ edited by\ \bibinfo {editor} {\bibfnamefont {G.~L.}\
  \bibnamefont {Bretthorst}}\ (\bibinfo  {publisher} {Cambridge University
  Press},\ \bibinfo {address} {Cambridge},\ \bibinfo {year} {2003})\BibitemShut
  {NoStop}%
\bibitem [{\citenamefont {Veitch}\ \emph {et~al.}(2015)\citenamefont {Veitch}
  \emph {et~al.}}]{veitch:2014wba}%
  \BibitemOpen
  \bibfield  {author} {\bibinfo {author} {\bibfnamefont {J.}~\bibnamefont
  {Veitch}} \emph {et~al.},\ }\href {\doibase 10.1103/PhysRevD.91.042003}
  {\bibfield  {journal} {\bibinfo  {journal} {Phys.Rev.}\ }\textbf {\bibinfo
  {volume} {D91}},\ \bibinfo {pages} {042003} (\bibinfo {year} {2015})},\
  \Eprint {http://arxiv.org/abs/1409.7215} {arXiv:1409.7215 [gr-qc]}
  \BibitemShut {NoStop}%
\bibitem [{\citenamefont {Cutler}\ and\ \citenamefont
  {Flanagan}(1994)}]{Cutler:1994ys}%
  \BibitemOpen
  \bibfield  {author} {\bibinfo {author} {\bibfnamefont {C.}~\bibnamefont
  {Cutler}}\ and\ \bibinfo {author} {\bibfnamefont {E.~E.}\ \bibnamefont
  {Flanagan}},\ }\href {\doibase 10.1103/PhysRevD.49.2658} {\bibfield
  {journal} {\bibinfo  {journal} {Phys. Rev.}\ }\textbf {\bibinfo {volume}
  {D49}},\ \bibinfo {pages} {2658} (\bibinfo {year} {1994})},\ \Eprint
  {http://arxiv.org/abs/gr-qc/9402014} {arXiv:gr-qc/9402014 [gr-qc]}
  \BibitemShut {NoStop}%
\bibitem [{\citenamefont {Abbott}\ \emph
  {et~al.}(2016{\natexlab{d}})\citenamefont {Abbott} \emph
  {et~al.}}]{GW150914-CALIBRATION}%
  \BibitemOpen
  \bibfield  {author} {\bibinfo {author} {\bibfnamefont {B.~P.}\ \bibnamefont
  {Abbott}} \emph {et~al.} (\bibinfo {collaboration} {LIGO Scientific
  Collaboration}),\ }\href@noop {} {\  (\bibinfo {year}
  {2016}{\natexlab{d}})},\ \bibinfo {note}
  {\url{https://dcc.ligo.org/LIGO-P1500248/public/main}},\ \Eprint
  {http://arxiv.org/abs/1602.03845} {arXiv:1602.03845 [gr-qc]} \BibitemShut
  {NoStop}%
\bibitem [{\citenamefont {Veitch}\ and\ \citenamefont
  {Del~Pozzo}(2013)}]{margphi}%
  \BibitemOpen
  \bibfield  {author} {\bibinfo {author} {\bibfnamefont {J.}~\bibnamefont
  {Veitch}}\ and\ \bibinfo {author} {\bibfnamefont {W.}~\bibnamefont
  {Del~Pozzo}},\ }\href {https://dcc.ligo.org/LIGO-T1300326} {\emph {\bibinfo
  {title} {{Analytic marginalisation of phase parameter}}}},\ \bibinfo {type}
  {Tech. Rep.}\ \bibinfo {number} {LIGO-T1300326}\ (\bibinfo  {institution}
  {LIGO Scientific Collaboration and Virgo Collaboration},\ \bibinfo {year}
  {2013})\BibitemShut {NoStop}%
\bibitem [{\citenamefont {Rubin}(1981)}]{rubin1981}%
  \BibitemOpen
  \bibfield  {author} {\bibinfo {author} {\bibfnamefont {D.~B.}\ \bibnamefont
  {Rubin}},\ }\href {\doibase 10.1214/aos/1176345338} {\bibfield  {journal}
  {\bibinfo  {journal} {Ann. Statist.}\ }\textbf {\bibinfo {volume} {9}},\
  \bibinfo {pages} {130} (\bibinfo {year} {1981})}\BibitemShut {NoStop}%
\bibitem [{\citenamefont {{Johnson-McDaniel}}\ \emph
  {et~al.}(2016)\citenamefont {{Johnson-McDaniel}} \emph
  {et~al.}}]{spinfit-T1600168}%
  \BibitemOpen
  \bibfield  {author} {\bibinfo {author} {\bibfnamefont {N.~K.}\ \bibnamefont
  {{Johnson-McDaniel}}} \emph {et~al.},\ }\href {https://dcc.ligo.org/T1600168}
  {\emph {\bibinfo {title} {Determining the final spin of a binary black hole
  system including in-plane spins: Method and checks of accuracy}}},\ \bibinfo
  {type} {Tech. Rep.}\ \bibinfo {number} {{LIGO}-T1600168}\ (\bibinfo
  {institution} {{LIGO} Project},\ \bibinfo {year} {2016})\BibitemShut
  {NoStop}%
\bibitem [{\citenamefont {Racine}(2008)}]{Racine:2008qv}%
  \BibitemOpen
  \bibfield  {author} {\bibinfo {author} {\bibfnamefont {E.}~\bibnamefont
  {Racine}},\ }\href {\doibase 10.1103/PhysRevD.78.044021} {\bibfield
  {journal} {\bibinfo  {journal} {Phys. Rev.}\ }\textbf {\bibinfo {volume}
  {D78}},\ \bibinfo {pages} {044021} (\bibinfo {year} {2008})},\ \Eprint
  {http://arxiv.org/abs/0803.1820} {arXiv:0803.1820 [gr-qc]} \BibitemShut
  {NoStop}%
\bibitem [{\citenamefont {Ajith}\ \emph {et~al.}(2011)\citenamefont {Ajith}
  \emph {et~al.}}]{Ajith:2009bn}%
  \BibitemOpen
  \bibfield  {author} {\bibinfo {author} {\bibfnamefont {P.}~\bibnamefont
  {Ajith}} \emph {et~al.},\ }\href {\doibase 10.1103/PhysRevLett.106.241101}
  {\bibfield  {journal} {\bibinfo  {journal} {Phys. Rev. Lett.}\ }\textbf
  {\bibinfo {volume} {106}},\ \bibinfo {pages} {241101} (\bibinfo {year}
  {2011})},\ \Eprint {http://arxiv.org/abs/0909.2867} {arXiv:0909.2867 [gr-qc]}
  \BibitemShut {NoStop}%
\bibitem [{\citenamefont {Santamar{\'i}a}\ \emph {et~al.}(2010)\citenamefont
  {Santamar{\'i}a}, \citenamefont {Ohme}, \citenamefont {Ajith}, \citenamefont
  {Br{\"u}gmann}, \citenamefont {Dorband}, \citenamefont {Hannam},
  \citenamefont {Husa}, \citenamefont {Moesta}, \citenamefont {Pollney},
  \citenamefont {Reisswig}, \citenamefont {Robinson}, \citenamefont {Seiler},\
  and\ \citenamefont {Krishnan}}]{Santamaria:2010}%
  \BibitemOpen
  \bibfield  {author} {\bibinfo {author} {\bibfnamefont {L.}~\bibnamefont
  {Santamar{\'i}a}}, \bibinfo {author} {\bibfnamefont {F.}~\bibnamefont
  {Ohme}}, \bibinfo {author} {\bibfnamefont {P.}~\bibnamefont {Ajith}},
  \bibinfo {author} {\bibfnamefont {B.}~\bibnamefont {Br{\"u}gmann}}, \bibinfo
  {author} {\bibfnamefont {N.}~\bibnamefont {Dorband}}, \bibinfo {author}
  {\bibfnamefont {M.}~\bibnamefont {Hannam}}, \bibinfo {author} {\bibfnamefont
  {S.}~\bibnamefont {Husa}}, \bibinfo {author} {\bibfnamefont {P.}~\bibnamefont
  {Moesta}}, \bibinfo {author} {\bibfnamefont {D.}~\bibnamefont {Pollney}},
  \bibinfo {author} {\bibfnamefont {C.}~\bibnamefont {Reisswig}}, \bibinfo
  {author} {\bibfnamefont {E.~L.}\ \bibnamefont {Robinson}}, \bibinfo {author}
  {\bibfnamefont {J.}~\bibnamefont {Seiler}}, \ and\ \bibinfo {author}
  {\bibfnamefont {B.}~\bibnamefont {Krishnan}},\ }\href {\doibase
  10.1103/PhysRevD.82.064016} {\bibfield  {journal} {\bibinfo  {journal}
  {\prd}\ }\textbf {\bibinfo {volume} {82}},\ \bibinfo {pages} {064016}
  (\bibinfo {year} {2010})},\ \Eprint {http://arxiv.org/abs/1005.3306}
  {arXiv:1005.3306 [gr-qc]} \BibitemShut {NoStop}%
\bibitem [{\citenamefont {Schmidt}\ and\ \citenamefont
  {Harry}(2016)}]{NRinjSchmidtHarryInPrep}%
  \BibitemOpen
  \bibfield  {author} {\bibinfo {author} {\bibfnamefont {P.}~\bibnamefont
  {Schmidt}}\ and\ \bibinfo {author} {\bibfnamefont {I.}~\bibnamefont
  {Harry}},\ }\href {https://dcc.ligo.org/LIGO-T1500606} {\enquote {\bibinfo
  {title} {{Numerical Relativity Injection Infrastructure}},}\ } (\bibinfo
  {year} {2016}),\ \bibinfo {note} {(in preparation)}\BibitemShut {NoStop}%
\bibitem [{\citenamefont {Schmidt}\ and\ \citenamefont
  {Galley}(2016)}]{NRSplineSchmidtGalleyInPrep}%
  \BibitemOpen
  \bibfield  {author} {\bibinfo {author} {\bibfnamefont {P.}~\bibnamefont
  {Schmidt}}\ and\ \bibinfo {author} {\bibfnamefont {C.}~\bibnamefont
  {Galley}},\ }\href {https://dcc.ligo.org/LIGO-P1600064} {\enquote {\bibinfo
  {title} {{Reduced-order spline interpolants of Numerical Relativity
  waveforms}},}\ } (\bibinfo {year} {2016}),\ \bibinfo {note} {(in
  preparation)}\BibitemShut {NoStop}%
\bibitem [{SpE()}]{SpECwebsite}%
  \BibitemOpen
  \href@noop {} {}\bibinfo {howpublished}
  {\url{http://www.black-holes.org/SpEC.html}}\BibitemShut {NoStop}%
\bibitem [{SXS()}]{SXSCatalog}%
  \BibitemOpen
  \href@noop {} {}\bibinfo {howpublished}
  {\url{http://www.black-holes.org/waveforms}}\BibitemShut {NoStop}%
\bibitem [{\citenamefont {Abbott}\ \emph
  {et~al.}(2016{\natexlab{e}})\citenamefont {Abbott} \emph
  {et~al.}}]{GW150914-RATES}%
  \BibitemOpen
  \bibfield  {author} {\bibinfo {author} {\bibfnamefont {B.~P.}\ \bibnamefont
  {Abbott}} \emph {et~al.} (\bibinfo {collaboration} {LIGO Scientific
  Collaboration, Virgo Collaboration}),\ }\href@noop {} {\  (\bibinfo {year}
  {2016}{\natexlab{e}})},\ \bibinfo {note}
  {\url{https://dcc.ligo.org/LIGO-P1500217/public/main}},\ \Eprint
  {http://arxiv.org/abs/1602.03842} {arXiv:1602.03842 [astro-ph.HE]}
  \BibitemShut {NoStop}%
\end{thebibliography}%

  \let\author\myauthor
  \let\affiliation\myaffiliation
  \let\maketitle\mymaketitle
  \title{Authors}
  \pacs{}
 



\author{%
B.~P.~Abbott,$^{1}$  
R.~Abbott,$^{1}$  
T.~D.~Abbott,$^{2}$  
M.~R.~Abernathy,$^{3}$  
F.~Acernese,$^{4,5}$ 
K.~Ackley,$^{6}$  
C.~Adams,$^{7}$  
T.~Adams,$^{8}$ 
P.~Addesso,$^{9}$  
R.~X.~Adhikari,$^{1}$
% ...
\\
\medskip
(LIGO Scientific Collaboration and Virgo Collaboration) 
\\
\medskip
{{}$^{\dag}$Deceased, May 2015. {}$^{\ddag}$Deceased, March 2015. }%
}\noaffiliation
\affiliation {LIGO, California Institute of Technology, Pasadena, CA 91125, USA }
\affiliation {Louisiana State University, Baton Rouge, LA 70803, USA }
\affiliation {American University, Washington, D.C. 20016, USA }
\affiliation {Universit\`a di Salerno, Fisciano, I-84084 Salerno, Italy }
\affiliation {INFN, Sezione di Napoli, Complesso Universitario di Monte S.Angelo, I-80126 Napoli, Italy }
\affiliation {University of Florida, Gainesville, FL 32611, USA }
\affiliation {LIGO Livingston Observatory, Livingston, LA 70754, USA }
\affiliation {Laboratoire d'Annecy-le-Vieux de Physique des Particules (LAPP), Universit\'e Savoie Mont Blanc, CNRS/IN2P3, F-74941 Annecy-le-Vieux, France }
\affiliation {University of Sannio at Benevento, I-82100 Benevento, Italy and INFN, Sezione di Napoli, I-80100 Napoli, Italy }
\affiliation {Albert-Einstein-Institut, Max-Planck-Institut f\"ur Gravi\-ta\-tions\-physik, D-30167 Hannover, Germany }
\affiliation {Nikhef, Science Park, 1098 XG Amsterdam, The Netherlands }
\affiliation {LIGO, Massachusetts Institute of Technology, Cambridge, MA 02139, USA }
\affiliation {Instituto Nacional de Pesquisas Espaciais, 12227-010 S\~{a}o Jos\'{e} dos Campos, S\~{a}o Paulo, Brazil }
\affiliation {INFN, Gran Sasso Science Institute, I-67100 L'Aquila, Italy }
\affiliation {INFN, Sezione di Roma Tor Vergata, I-00133 Roma, Italy }
\affiliation {Inter-University Centre for Astronomy and Astrophysics, Pune 411007, India }
\affiliation {International Centre for Theoretical Sciences, Tata Institute of Fundamental Research, Bangalore 560012, India }
\affiliation {University of Wisconsin-Milwaukee, Milwaukee, WI 53201, USA }
\affiliation {Leibniz Universit\"at Hannover, D-30167 Hannover, Germany }
\affiliation {Universit\`a di Pisa, I-56127 Pisa, Italy }
\affiliation {INFN, Sezione di Pisa, I-56127 Pisa, Italy }
\affiliation {Australian National University, Canberra, Australian Capital Territory 0200, Australia }
\affiliation {The University of Mississippi, University, MS 38677, USA }
\affiliation {California State University Fullerton, Fullerton, CA 92831, USA }
\affiliation {LAL, Univ. Paris-Sud, CNRS/IN2P3, Universit\'e Paris-Saclay, Orsay, France }
\affiliation {Chennai Mathematical Institute, Chennai 603103, India }
\affiliation {Universit\`a di Roma Tor Vergata, I-00133 Roma, Italy }
\affiliation {University of Southampton, Southampton SO17 1BJ, United Kingdom }
\affiliation {Universit\"at Hamburg, D-22761 Hamburg, Germany }
\affiliation {INFN, Sezione di Roma, I-00185 Roma, Italy }
\affiliation {Albert-Einstein-Institut, Max-Planck-Institut f\"ur Gravitations\-physik, D-14476 Potsdam-Golm, Germany }
\affiliation {APC, AstroParticule et Cosmologie, Universit\'e Paris Diderot, CNRS/IN2P3, CEA/Irfu, Observatoire de Paris, Sorbonne Paris Cit\'e, F-75205 Paris Cedex 13, France }
\affiliation {Montana State University, Bozeman, MT 59717, USA }
\affiliation {Universit\`a di Perugia, I-06123 Perugia, Italy }
\affiliation {INFN, Sezione di Perugia, I-06123 Perugia, Italy }
\affiliation {European Gravitational Observatory (EGO), I-56021 Cascina, Pisa, Italy }
\affiliation {Syracuse University, Syracuse, NY 13244, USA }
\affiliation {SUPA, University of Glasgow, Glasgow G12 8QQ, United Kingdom }
\affiliation {LIGO Hanford Observatory, Richland, WA 99352, USA }
\affiliation {Wigner RCP, RMKI, H-1121 Budapest, Konkoly Thege Mikl\'os \'ut 29-33, Hungary }
\affiliation {Columbia University, New York, NY 10027, USA }
\affiliation {Stanford University, Stanford, CA 94305, USA }
\affiliation {Universit\`a di Padova, Dipartimento di Fisica e Astronomia, I-35131 Padova, Italy }
\affiliation {INFN, Sezione di Padova, I-35131 Padova, Italy }
\affiliation {CAMK-PAN, 00-716 Warsaw, Poland }
\affiliation {University of Birmingham, Birmingham B15 2TT, United Kingdom }
\affiliation {Universit\`a degli Studi di Genova, I-16146 Genova, Italy }
\affiliation {INFN, Sezione di Genova, I-16146 Genova, Italy }
\affiliation {RRCAT, Indore MP 452013, India }
\affiliation {Faculty of Physics, Lomonosov Moscow State University, Moscow 119991, Russia }
\affiliation {SUPA, University of the West of Scotland, Paisley PA1 2BE, United Kingdom }
\affiliation {University of Western Australia, Crawley, Western Australia 6009, Australia }
\affiliation {Department of Astrophysics/IMAPP, Radboud University Nijmegen, P.O. Box 9010, 6500 GL Nijmegen, The Netherlands }
\affiliation {Artemis, Universit\'e C\^ote d'Azur, CNRS, Observatoire C\^ote d'Azur, CS 34229, Nice cedex 4, France }
\affiliation {Institut de Physique de Rennes, CNRS, Universit\'e de Rennes 1, F-35042 Rennes, France }
\affiliation {Washington State University, Pullman, WA 99164, USA }
\affiliation {Universit\`a degli Studi di Urbino ``Carlo Bo,'' I-61029 Urbino, Italy }
\affiliation {INFN, Sezione di Firenze, I-50019 Sesto Fiorentino, Firenze, Italy }
\affiliation {University of Oregon, Eugene, OR 97403, USA }
\affiliation {Laboratoire Kastler Brossel, UPMC-Sorbonne Universit\'es, CNRS, ENS-PSL Research University, Coll\`ege de France, F-75005 Paris, France }
\affiliation {Carleton College, Northfield, MN 55057, USA }
\affiliation {Astronomical Observatory Warsaw University, 00-478 Warsaw, Poland }
\affiliation {VU University Amsterdam, 1081 HV Amsterdam, The Netherlands }
\affiliation {University of Maryland, College Park, MD 20742, USA }
\affiliation {Center for Relativistic Astrophysics and School of Physics, Georgia Institute of Technology, Atlanta, GA 30332, USA }
\affiliation {Laboratoire des Mat\'eriaux Avanc\'es (LMA), CNRS/IN2P3, F-69622 Villeurbanne, France }
\affiliation {Universit\'e Claude Bernard Lyon 1, F-69622 Villeurbanne, France }
\affiliation {Universit\`a di Napoli ``Federico II,'' Complesso Universitario di Monte S.Angelo, I-80126 Napoli, Italy }
\affiliation {NASA/Goddard Space Flight Center, Greenbelt, MD 20771, USA }
\affiliation {RESCEU, University of Tokyo, Tokyo, 113-0033, Japan. }
\affiliation {Tsinghua University, Beijing 100084, China }
\affiliation {Texas Tech University, Lubbock, TX 79409, USA }
\affiliation {National Tsing Hua University, Hsinchu City, 30013 Taiwan, Republic of China }
\affiliation {Charles Sturt University, Wagga Wagga, New South Wales 2678, Australia }
\affiliation {West Virginia University, Morgantown, WV 26506, USA }
\affiliation {University of Chicago, Chicago, IL 60637, USA }
\affiliation {Caltech CaRT, Pasadena, CA 91125, USA }
\affiliation {Korea Institute of Science and Technology Information, Daejeon 305-806, Korea }
\affiliation {Universit\`a di Roma ``La Sapienza,'' I-00185 Roma, Italy }
\affiliation {University of Brussels, Brussels 1050, Belgium }
\affiliation {Sonoma State University, Rohnert Park, CA 94928, USA }
\affiliation {Center for Interdisciplinary Exploration \& Research in Astrophysics (CIERA), Northwestern University, Evanston, IL 60208, USA }
\affiliation {University of Minnesota, Minneapolis, MN 55455, USA }
\affiliation {The University of Melbourne, Parkville, Victoria 3010, Australia }
\affiliation {Institute for Plasma Research, Bhat, Gandhinagar 382428, India }
\affiliation {The University of Sheffield, Sheffield S10 2TN, United Kingdom }
\affiliation {The University of Texas Rio Grande Valley, Brownsville, TX 78520, USA }
\affiliation {Universit\`a di Trento, Dipartimento di Fisica, I-38123 Povo, Trento, Italy }
\affiliation {INFN, Trento Institute for Fundamental Physics and Applications, I-38123 Povo, Trento, Italy }
\affiliation {The Pennsylvania State University, University Park, PA 16802, USA }
\affiliation {Cardiff University, Cardiff CF24 3AA, United Kingdom }
\affiliation {Montclair State University, Montclair, NJ 07043, USA }
\affiliation {MTA E\"otv\"os University, ``Lendulet'' Astrophysics Research Group, Budapest 1117, Hungary }
\affiliation {National Astronomical Observatory of Japan, 2-21-1 Osawa, Mitaka, Tokyo 181-8588, Japan }
\affiliation {School of Mathematics, University of Edinburgh, Edinburgh EH9 3FD, United Kingdom }
\affiliation {Indian Institute of Technology, Gandhinagar Ahmedabad Gujarat 382424, India }
\affiliation {University of Szeged, D\'om t\'er 9, Szeged 6720, Hungary }
\affiliation {Embry-Riddle Aeronautical University, Prescott, AZ 86301, USA }
\affiliation {Tata Institute of Fundamental Research, Mumbai 400005, India }
\affiliation {INAF, Osservatorio Astronomico di Capodimonte, I-80131, Napoli, Italy }
\affiliation {University of Michigan, Ann Arbor, MI 48109, USA }
\affiliation {Rochester Institute of Technology, Rochester, NY 14623, USA }
\affiliation {NCSA, University of Illinois at Urbana-Champaign, Urbana, Illinois 61801, USA }
\affiliation {Universitat de les Illes Balears, IAC3---IEEC, E-07122 Palma de Mallorca, Spain }
\affiliation {University of Bia{\l }ystok, 15-424 Bia{\l }ystok, Poland }
\affiliation {SUPA, University of Strathclyde, Glasgow G1 1XQ, United Kingdom }
\affiliation {IISER-TVM, CET Campus, Trivandrum Kerala 695016, India }
\affiliation {Canadian Institute for Theoretical Astrophysics, University of Toronto, Toronto, Ontario M5S 3H8, Canada }
\affiliation {Institute of Applied Physics, Nizhny Novgorod, 603950, Russia }
\affiliation {Pusan National University, Busan 609-735, Korea }
\affiliation {Hanyang University, Seoul 133-791, Korea }
\affiliation {University of Adelaide, Adelaide, South Australia 5005, Australia }
\affiliation {NCBJ, 05-400 \'Swierk-Otwock, Poland }
\affiliation {IM-PAN, 00-956 Warsaw, Poland }
\affiliation {Monash University, Victoria 3800, Australia }
\affiliation {Seoul National University, Seoul 151-742, Korea }
\affiliation {The Chinese University of Hong Kong, Shatin, NT, Hong Kong SAR, China }
\affiliation {University of Alabama in Huntsville, Huntsville, AL 35899, USA }
\affiliation {University of Massachusetts-Amherst, Amherst, MA 01003, USA }
\affiliation {ESPCI, CNRS, F-75005 Paris, France }
\affiliation {Universit\`a di Camerino, Dipartimento di Fisica, I-62032 Camerino, Italy }
\affiliation {Southern University and A\&M College, Baton Rouge, LA 70813, USA }
\affiliation {College of William and Mary, Williamsburg, VA 23187, USA }
\affiliation {Instituto de F\'\i sica Te\'orica, University Estadual Paulista/ICTP South American Institute for Fundamental Research, S\~ao Paulo SP 01140-070, Brazil }
\affiliation {University of Cambridge, Cambridge CB2 1TN, United Kingdom }
\affiliation {IISER-Kolkata, Mohanpur, West Bengal 741252, India }
\affiliation {Rutherford Appleton Laboratory, HSIC, Chilton, Didcot, Oxon OX11 0QX, United Kingdom }
\affiliation {Whitman College, 345 Boyer Avenue, Walla Walla, WA 99362 USA }
\affiliation {National Institute for Mathematical Sciences, Daejeon 305-390, Korea }
\affiliation {Universit\'e de Lyon, F-69361 Lyon, France }
\affiliation {Hobart and William Smith Colleges, Geneva, NY 14456, USA }
\affiliation {Janusz Gil Institute of Astronomy, University of Zielona G\'ora, 65-265 Zielona G\'ora, Poland }
\affiliation {King's College London, University of London, London WC2R 2LS, United Kingdom }
\affiliation {Andrews University, Berrien Springs, MI 49104, USA }
\affiliation {Universit\`a di Siena, I-53100 Siena, Italy }
\affiliation {Trinity University, San Antonio, TX 78212, USA }
\affiliation {University of Washington, Seattle, WA 98195, USA }
\affiliation {Kenyon College, Gambier, OH 43022, USA }
\affiliation {Abilene Christian University, Abilene, TX 79699, USA }
\affiliation {Cornell University, Ithaca, NY 14853, USA }
\affiliation {Theoretical Physics Institute, University of Jena, 07743 Jena, Germany }
\affiliation {Caltech JPL, Pasadena, CA 91109, USA }


  \newpage
  \maketitle

\end{document}